\def\O5{\Omega_5}

\def\Yo{{Y}}
\def\Ym2{{_{-2}Y}}
\def\Yp2{{_{2}Y}}

\def\P{{\bf P}}

\def\ellmin{\ell_{\rm min}}
\def\ellmax{\ell_{\rm max}}
\def\fsky{f_{\rm sky}}
\def\simle{\lesssim}
\def\simge{\gtrsim}
\def\lm{{lm}}
\def\elm{{E,lm}}
\def\blm{{B,lm}}

\def\tlm{{2,lm}}
\def\mtlm{{-2,lm}}
\def\lm{{lm}}

\def\W{{\bf W}}

\def\l{{\bf l}}

\def\k{{\bf l}}

\def\r{{\bf x}}

\def\beq{\begin{equation}}
\def\eeq{\end{equation}}
\def\beqa{\begin{eqnarray}}
\def\eeqa{\end{eqnarray}}

\newcommand{\ra}{\;\raise1.0pt\hbox{$'$}\hskip-6pt\partial\;}
\newcommand{\lo}{\;\overline{\raise1.0pt\hbox{$'$}\hskip-6pt
\partial}\;}

\def\be{\begin{equation}}
\def\ee{\end{equation}}
\def\ba{\begin{eqnarray}}
\def\ea{\end{eqnarray}}
\def\nn{\nonumber}
\def\eqdef{\stackrel{\rm def}{=}}
\def\one{{\bf 1}_{\rm vec}}
\def\bC{{\bf C}}
\def\bCa{{\bf C}^\alpha}
\def\bd{{\bf d}}
\def\bp{{\bf \Phi}^{\alpha}}
\def\tO{{\mathcal O}}
\def\npix{N_{\rm pix}}
\def\Wopt{{\bf W}_{\rm opt}}

\def\ta{{\widetilde a}}
\def\tC{{\widetilde C}}
\def\hC{{\widehat C}}
\def\bx{{\bf x}}
\def\cc{\mbox{c.c.}}
\def\Cov{\mbox{Cov}}

\def\qfive{\qquad\qquad\qquad\qquad\qquad}
\def\tW{\widetilde W}
\def\eE{\langle \tC_\alpha \rangle}
\def\eEs{\langle \tC_\alpha \rangle_{\rm sig}}
\def\eEn{\langle \tC_\alpha \rangle_{\rm noise}}
\def\threej#1#2#3#4#5#6{\left( \begin{array}{ccc} #1 & #2 & #3 \\ #4 & #5 & #6 \end{array} \right) }
\def\bigoh{{\mathcal O}}

\documentclass[prd,aps,epsf,superscriptaddress]{revtex4}
\usepackage{epsfig}

\begin{document}

\title{A general solution to the $E$-$B$ mixing problem}
\author{Kendrick M. Smith}
\affiliation{Kavli Institute for Cosmological Physics, Enrico Fermi Institute, University of Chicago, 60637}
\affiliation{Department of Physics,  University of Chicago, 60637}
\author{Matias Zaldarriaga}
\affiliation{Center for Astrophysics, Harvard University, 02138}
\affiliation{Jefferson Physical Laboratory, Harvard University, 02138}

\vskip 1pc


\begin{abstract}
We derive a general ansatz for optimizing pseudo-$C_\ell$ estimators used to measure CMB anisotropy power spectra,
and apply it to the recently-proposed pure pseudo-$C_\ell$ formalism, to obtain an estimator which achieves near-optimal
$B$-mode power spectrum errors for any specified noise distribution while minimizing leakage from ambiguous modes.
Our technique should be relevant for upcoming CMB polarization experiments searching for $B$-mode polarization. 
We compare our technique both to the theoretical limits based on a full Fisher matrix calculation and to the standard pseudo-$C_\ell$ technique.
We demonstrate it by applying it to a fiducial survey with realistic inhomogeneous noise, complex boundaries, point source masking, and
noise level comparable to what is expected for next generation experiments ($\sim$5.75 $\mu$K-arcmin).
For such an experiment our technique could improve the constraints on the amplitude of a gravity wave background by over a factor of ten 
compared to what could be obtained using ordinary pseudo-$C_\ell$, coming quite close to saturating the theoretical limit. 
Constraints on the amplitude of the lensing $B$-modes are improved by about a factor of 3. 
\end{abstract}

\maketitle

\section{Introduction}

The polarization of the Cosmic Microwave Background (CMB) could provide a unique window into the very early Universe if its  pattern can decomposed into  gradient and curl (or $E$ and $B$) parts. The detection of large scale $B$-mode polarization would indicate the presence of a stochastic background of gravitational waves left over from the epoch of Inflation \cite{spinlett,kkslett}. The amplitude of the background would allow us to directly deduce the energy scale of Inflation, dramatically extending our understanding of the history of  Universe to its very beginnings. $B$-mode polarization on small angular scales is mainly produced by the gravitational lensing of the $E$-mode component \cite{pollens}. A detailed measure of these contributions could help us constrain various cosmological parameters such as the equation of state  of the dark energy  or the mass of the neutrinos. 

The clear prospect of studying the birth of our Universe and improving constraints on some cosmological parameters using the CMB polarization has led to  remarkable progress in detector technology with the promise to deliver the required sensitivity to detect the small $B$ polarization signal. In the near future experiments targeting $B$-mode polarization will be performed from the ground or from balloons. These experiments will cover relatively small fractions of the sky with the aim of producing high fidelity polarization maps that can be decomposed into $E$ and $B$.

In a finite patch of sky, the $E$-$B$ decomposition framework needs  modification. It is no longer true that any polarization field can be uniquely decomposed into pure $E$ and pure $B$-modes. A new set of modes, so-called ``ambiguous'', needs to be introduced \cite{Bunn:2002df}. These new modes receive contributions from both $E$ and $B$-modes. In practice because the $E$ signal is expected to be so much larger than the $B$ one, the ambiguous modes will be dominated by $E$ signal. Thus all the information about the cosmological $B$-modes is contained in the pure $B$-modes, which are orthogonal to both pure $E$ and ambiguous modes. The leakage of $E$ signal into $B$ can be thought of as an additional source of noise on top of that introduced by the detector, but a source of noise that can be eliminated by a suitable choice of variables. 

Analyzing the data of the next generation of experiments is potentially quite challenging. Measuring the power spectra using a fully optimal likelihood based analysis requires $O(N_{pix}^3)$ operations and is probably not feasible  for this next generation of experiments. The alternative technique, widely used for CMB temperature experiments, is the pseudo-$C_\ell$ quadratic method \cite{Wandelt:1998qd}. Unfortunately in a finite patch of the sky the pseudo-$C_\ell$'s do not isolate the pure $B$-modes before constructing the quadratic estimators. As a result the quadratic estimator is contamined by the leakage from the ambiguous modes. This contamination can be removed on average by taking suitable linear combinations of the pseudo-$C_\ell$'s. However the leaked power still contributes to the variance of the estimators, significantly degrading their ability to measure the amplitude of the cosmological $B$-modes. 

Recently an improved technique, a pure pseudo-$C_\ell$ technique was introduced \cite{Smith:2005gi}. This technique preserves the simplicity of the pseudo-$C_\ell$'s  but at the same time ensures that no $E \rightarrow B$ mixing occurs.  In \cite{Smith:2005gi} it was also shown that the pure pseudo-$C_\ell$'s  preserve most of the cosmological information in the data. 

In this paper we will look at the pure pseudo-$C_\ell$ from a new perspective.  
Our different approach will allow us to understand various features of the pure pseudo-$C_\ell$'s and also give 
guidance as to how to choose the weight function that the technique requires.
The main result of the paper, a general ansatz for optimizing the pseudo-$C_\ell$ weight function given arbitrary signal and noise covariance,
is presented in \S\ref{sec:conjecture}.
Using this ansatz to generate weight functions in an automated way, we study several mock surveys, culminating in a realistic
simulation of a fiducial experiment, with characteristics based on the upcoming balloon based EBEX \cite{Oxley:2005dg}. Our simulations have inhomogeneous noise, complicated boundaries and point source masking. We will argue that the pseudo-$C_\ell$ estimators give near-optimal 
power spectrum errors on all angular scales.
This resolves a long-standing practical issue in the pseudo-$C_\ell$ method: the lack of an algorithm for choosing weight functions in complex situations.
Previous studies have had to rely on optimizing weight functions by hand, using expensive Monte Carlo simulations to compute
the estimator variance for each candidate weight function considered.
At the same time, we preserve the strong $E$-$B$ separation of the pure pseudo-$C_\ell$ method, achieving excellent $B$-mode errors
even for the low noise levels anticipated for the upcoming generation of CMB polarization experiments \cite{Oxley:2005dg,QUIET,Taylor:2004hh,Yoon:2006jc}.

\section{Preliminaries}

\subsection{Spin two notation}

First we will introduce our notation. We will follow closely \cite{Bunn:2002df}.
The (linear) polarization of the CMB is described in terms of the Stokes
parameters $Q$ and $U$. The definition of $Q$ and $U$ depends on the
coordinate system chosen. In this subsection we review definitions
that are valid
for the full sky, so we will use spherical coordinates to
define $Q$ and $U$. 

We will follow the notation of \cite{3.spinlong}.  The Stokes parameters can
be combined to form a spin-2 combination $(Q+iU)$ and a spin-$(-2)$ combination $(Q-iU)$.
In the full sky these combinations can be decomposed using spin-2 harmonics, 
\beqa Q+iU= \sum_\lm a_\tlm \ \Yp2_\lm \ \ \ &;&\ \
\ Q-iU= \sum_\lm a_\mtlm \ \Ym2_\lm
\label{decomp}
\eeqa

It is natural to introduce a scalar ($E$) and a pseudoscalar ($B$) 
field to describe polarization. The expansion coefficients of these
two fields in (ordinary spin-0) spherical harmonics are
\beqa
a_\elm= -(a_\tlm+a_\mtlm)/2 \ \ \ &;& \ \ \
a_\blm= i(a_\tlm-a_\mtlm)/2 .
\label{eb}
\eeqa
These two functions completely characterize any
polarization field on the sphere \cite{3.spinlong}. 
They are important physically because
cosmological density perturbations cannot create $B$-type polarization
while gravity waves can \cite{2.kks,3.spinlong}. 

The spin-2 harmonics in equation (\ref{decomp}) can be related to the
usual spin-0 spherical harmonics by means of two first-order differential
operators, the spin-raising ($\ra$) and
spin-lowering ($\lo$) operators \cite{3.spinlong}.
When applied to the spin-weighted spherical harmonics, these operators
yield the following identities:
\beqa
\ra _{s}\Yo_\lm &=& [(l-s)(l+s+1)]^{1/2}\ _{s+1}\Yo_\lm \nonumber \\
\lo _{s}\Yo_\lm &=& -[(l+s)(l-s+1)]^{1/2}\ _{s-1}\Yo_\lm.
\label{ypym}
\eeqa 

We can  show that:
\beqa 
\label{chidef}
\chi_E &\equiv& [\lo\lo(Q+iU) + \ra\ra (Q-iU)]/2 \nonumber \\ 
&=& - \sum_\lm [(l+2)!/(l-2)!]^{1/2} a_\elm \Yo_\lm  \nonumber \\
\chi_B &\equiv& i [\lo\lo(Q+iU) - \ra\ra (Q-iU)]/2 \nonumber \\
 &=&  \sum_\lm [(l+2)!/(l-2)!]^{1/2} a_\blm \Yo_\lm.  \label{eq:chiB}
\eeqa 
Thus we can take linear combinations of second
derivatives of the Stokes parameters
and obtain variables that depend only on $E$
or on $B$.  

We pause to note that the reason why $E$ and $B$ are the focus of
attention instead of  $\chi_E$,  $\chi_B$  is partly a matter of
convention. Perhaps more importantly, $E$ and $B$ have the same
power spectrum on small scales as the Stokes parameters, while that of $\chi$ differs by a factor
of $\sim l^{4}$. As a result while white noise in $Q$ and $U$ translates into white noise in $E$ and $B$, it becomes colored noise for $\chi$.

\subsection{Small-angle approximation}
\label{flatsky}

If one works over a small patch of sky, one can use  the 
small-angle (flat-sky) approximation. When working in this limit,
it is more natural to measure the Stokes
parameters with respect to a Cartesian coordinate system $(x,y)$
instead of the usual polar coordinate axis. 
In the flat-sky approximation polarization is expanded in terms of Fourier modes,
\beqa
\left(\matrix{Q(\r) \cr U(\r)} \right) &=&  \int {d^2 l \over (2\pi)^2}
\left[E(\k ) \left(\matrix{\cos 2\phi  \cr \sin 2\phi}
\right) + B(\k ) \left(\matrix{-\sin 2\phi  \cr \cos 2\phi}
\right)\right]
e^{i\k\cdot\r},
\label{fourier}
\eeqa
where $\r=(x, y)$ and $\k=l (\cos\phi,\sin\phi)$.
The differential operators reduce to simply
\begin{eqnarray}
\ra &=& -(\partial_x+i\partial_y),\\
\lo &=& -(\partial_x-i\partial_y).
\label{DBdefEq}
\end{eqnarray}

\subsection{The ambiguous modes}

On a manifold without boundary, any polarization field
can be uniquely separated into an $E$ part and a $B$ part.  But
if there is a boundary ({\it i.e.}, if only
some subset $\Omega$ of the sky has been observed), 
this decomposition is not unique. In this section we summarize the results of \cite{Bunn:2002df}.

Polarization fields living on $\Omega$ form a normed vector space with the inner product 
\beq\label{InnerProductEq}
\int_\Omega (QQ' + U U')  d\Omega,
\eeq
and we say that two fields $(Q+iU)$ and $(Q'+iU')$ are orthogonal if their inner product vanishes. 
We refer to a polarization field as 
\begin{itemize}
\item $E$ if it has vanishing $\chi_B$,
\item $B$ if it has vanishing $\chi_E$,
\item pure $E$ if it is orthogonal to all $B$-fields, and
\item pure $B$ if it is orthogonal to all $E$-fields.
\end{itemize}

On the complete sky, every polarization field can be uniquely represented
as a linear combination of an $E$ field and a $B$ field, and all $E$ fields
are perpendicular to all $B$ fields.  In other words, the space of all
polarization fields is the direct sum of two orthogonal subspaces: 
the space of all
$E$ fields and the space of all $B$ fields.  
In this case, there is no distinction between an $E$ field
and a ``pure $E$'' field.

But if only some subset of the sky has been observed, so that $\Omega$ is a manifold
with boundary, then
this decomposition is not unique.
One way to see this is to note that there are modes that satisfy
both the $E$-mode and $B$-mode conditions simultaneously.
When we split a polarization field into an $E$ part and a $B$ part,
these ``ambiguous'' modes can go into either component.
In order to make the $E/B$ decomposition unique, we must first
project out the ambiguous modes.  

The existence of ambiguous modes has important implications at the time of measuring the $B$-mode power spectrum. Here we concentrate on ``simple''  quadratic methods that make no attempt to filter those ambiguous modes at the level of the map (such as \cite{Wandelt:1998qd}). A full likelihood analysis would automatically deal with the ambiguous modes but it will probably be computationally prohibitive. It might be feasible for low resolution maps,  becoming the method of choice for constraining the gravitational wave amplitude \cite{Efstathiou:2006eb}.

\begin{figure}
\centerline{\epsfxsize=4.0truein\epsffile{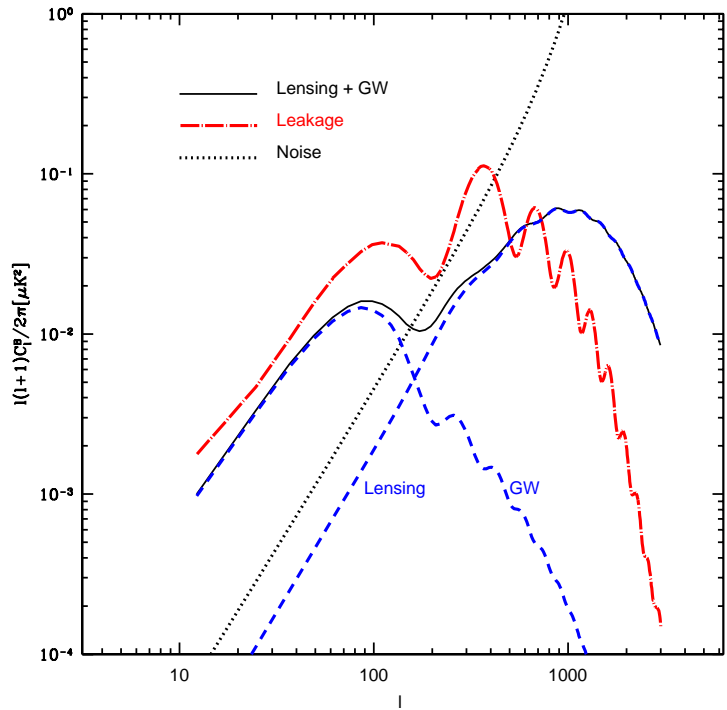}}
\caption{Illustration of the leakage due to finite sky coverage in the ordinary pseudo-$C_\ell$ technique. A circular region 10 degrees in radius was used to estimate the pseudo-$C_\ell$'s. The leakage curve shows the $B$-modes obtained even in the absence of any cosmological $B$-modes just from the leakage of $E$. For comparison we also show the expected cosmological signal for $T/S=0.1$ and the noise power spectrum for an experiment with $w_P^{-1/2}=5.75$ $\mu$K-arcmin and an 8 arcminute beam. }
\label{figleak}
\end{figure}

It is easiest to understand the issue to focus on the pseudo-$C_\ell$ method, but the point applies generally. 
In the pseudo-$C_\ell$ method one just measures the power spectra of $E$ and $B$ as one would do in the full sky and simply masks out the unobserved regions. After this procedure both $C_\ell^E$ and $C_\ell^B$ spectra are dominated by $E$-modes. An illustrative example is shown in figure \ref{figleak}. 
We considered a circular region $10^\circ$ in radius. The $E$ and $B$ power spectra were measured by simply masking the unobserved region but otherwise proceeding as if one had a full sky measurement.
The calculation was performed in the flat sky approximation which is adequate for this size region.
The input polarization field had $E$-modes only. In the figure we show the measured $B$  mode power spectrum spectra, which is produced by leakage of the $E$-modes. For comparison  we show the expected $B$-mode power spectrum if $T/S=0.1$ together with the $B$-modes produced by gravitational lensing. Also for reference we plot the detector noise power spectrum for $w_P^{-1/2}= 5.75$ $\mu$K-arcmin and an 8 arcminute beam. This level of noise is within the range of expected values for the upcoming EBEX balloon experiment \cite{Oxley:2005dg}.

The pseudo-$C_\ell$ method then takes linear combinations of the measured $C_\ell$'s to eliminate the $E$-mode contribution in the hopes of revealing the cosmological $B$-modes. The point is that this procedure makes the $B$-mode power spectra estimator independent of $E$ in the mean, but the variance of the estimator is still dominated by the leaked power. 

Perhaps an extreme example is helpful to clarify the problem further. Imagine that cosmological parameters were known perfectly so that there is no uncertainty in the  $E$-mode spectra. Then the leaked $B$
power is perfectly predicted and can be subtracted from the measured signal to leave an unbiased $C^B_\ell$ estimator.
This is analogous to what one does with detector noise. Unfortunately just as with noise the variance of the estimator is increased by the leaked power. 
The pseudo-$C_\ell$ technique does even worse than this because it does not assume any knowledge of the $C^E_\ell$ so it uses only linear combinations that are on average independent of $E$ for any choice of $C^E_\ell$.
(In reality for a small patch this is not possible and some assumptions such as constancy of the $E$ power spectrum in bands must be used. 
These assumptions are expected to be reasonably well satisfied.)

Figure \ref{figleak} illustrates the severity of the problem for experiments with noise levels comparable to those of our fiducial experiment (5.75 $\mu$K-arcmin). The leaked power is larger than the detector noise all the way to $l\sim 400$, so using the pseudo-$C_\ell$ technique is clearly wasteful. Note that even on scales where  the lensing signal dominates there would be significant  degradation.

\section{Avoiding leakage by measuring $\chi_B$}

When trying to measure the $B$-mode power spectrum in a finite patch 
there is a simple way to avoid unwanted leakage from $E$-modes: estimate 
the power spectrum starting from the $\chi_B$ map. In the next section we will show that this is mathematically equivalent to the pure pseudo-$C_\ell$ technique. It is important to realize that in practice 
no information is being lost by doing this. By construction $\chi_B$ has no contribution from either pure $E$-modes or ambiguous modes. Although in principle there is some information on the amplitude of $B$-modes in the ambiguous modes, in practice that signal is completely swamped by the contribution of the $E$-modes so there is no real loss of information by working with $\chi_B$. 

Perhaps an analogy  helps at this point. One could consider a scalar field, like the CMB temperature, and wonder whether there is any loss of information if one were to compute the Laplacian of the map and extract the power spectrum from that. In the full sky there would be no loss of information but in a small patch one would lose. In the full sky there is a one to one relation between the Laplacian map and the underlying temperature map
(except for the monopole and dipole which are lost when one takes the Laplacian, but they do not carry cosmological information);
the only difference is that both signal and noise power spectra will be multiplied by $\ell^2(\ell+1)^2$ which does not alter how well one can constrain each multipole. 
In a finite patch the situation is different because there are non-zero temperature fields on the finite patch  that can have zero Laplacian. 
Those modes are filtered out by the ``Laplacian-technique''  with the resulting loss of information.   
In our case the analogue of the modes that are being filtered out are the ambiguous modes which we want to remove anyway. 
Thus for constraining the $B$-mode amplitude, starting with the $\chi_B$ field is sufficient. 

Of course the fact that the $\chi_B$ map contains for practical purposes all the cosmological information does not mean that if we extract its power spectrum in a non-optimal way we will still conserve all the information in the data. If in practice (as we will do in later sections) we use a pseudo-$C_\ell$ technique to extract power spectra, it may be that allowing some leakage is still advantageous. As long as the leaked power is small compared to the detector noise it is most certainly harmless. We will see that it is indeed the case that small amounts of leakage are preferred by our technique. We will postpone discussion of this point to later. 

Once we conclude that measuring the power spectrum starting from $\chi_B$ avoids the unwanted leakage there are various technical issues one needs to address to implement such a method in practice. Moreover, we will show that this technique is mathematically equivalent to the pure pseudo-$C_\ell$ technique and we will find it easier to implement our algorithms using that perspective. We do want to comment on two aspects;  first, in practice one needs to take the derivatives by finite differencing which will introduce some error and residual leakage. This residual leakage will constrain the choice of pixelization. We discuss this in more detail in Appendix~\ref{finite}.
Second, the $\chi_B$ field has a very blue spectrum which will affect standard estimation techniques. We will discuss this second issue here because it will illuminate our discussions down the line.

\subsection{A very blue spectrum: the need for apodization}
\label{ssec:apodization}

The additional derivatives needed to construct $\chi_B$ make its power spectrum very blue. 
We will show in this section that this induces high levels of aliasing unless appropriate apodizations are used.  Thus apodization will be an integral part of any technique that starts with $\chi_B$. 

To understand the issue it is best to write down the formulas in the flat sky approximation. 
We assume that the Stokes parameters have been measured in a certain region of the sky  and that $\chi_B$ was calculated by finite differencing. Then the $\chi_B$ field is multiplied by a window function $W$ (in real space) and the Fourier components of $(W \chi_B)$ are measured and squared to estimate the power spectrum. The weight $W$ is zero outside the observed patch. If derivatives are taken using nearest neighbors, one also loses an additional pixel around the edge of the map. Note that this loss is directly related to the counting of ambiguous modes as described in \cite{Bunn:2002df}. 

The resulting power estimates are the convolution of the $\chi_B$ power spectrum and the Fourier transform of the window,
\beq
\hat C_\l^{\chi_B} = \int {d^2 l^\prime \over (2 \pi)^2} || \tW(\l - \l^{\prime})||^2 \ C_{l^\prime}^{\chi_B}.
\label{cl}
\eeq 
The $\chi_B$ power spectrum is now so blue that unless the window function decreases fast  with $l$, all estimates of power will be dominated by the small scale modes. 

One way of seeing why this is a problem is to consider normalizing the window such that $|| \tW(0)||^2=1$. The estimator in equation (\ref{cl}) estimates the power at $\l$ but has aliasing from other modes. Even if one considered the idealized situation in which the higher $l$ power spectra were known so that the aliasing could be calculated and subtracted,  the aliased power would still contribute to the noise of the estimator. If the aliased power dominates the integral in Eq.~(\ref{cl}), the additional variance is large compared to the signal one is after with the resulting loss of information. Thus one would want to avoid having such aliasing if possible. 

The high $l$ behavior of $|| \tW(\l )||^2$ is directly related to the number of normal derivatives of $W$ that vanish at the boundary of the patch. In fact 
\beq
\lim_{l \rightarrow \infty}  || \tW(\l )||^2 \propto l^{-2(n+2)},   \label{eq:nderivs}
\eeq
where $n$ is the number of derivatives of the window that vanish at the boundary with $n=0$ meaning that the window itself is continuous. Thus requiring that equation (\ref{cl}) converges requires using a window that is smoother at the boundary. For example if we compare measuring the power spectra of the Stokes parameters without apodization to measuring the power spectrum of $\chi_B$,  compensating the additional $l^4$ factor in the $\chi_B$ spectrum requires having both the window and its first derivative vanish. 

Reducing the aliasing requires considering weights that decrease continuously to zero as one gets to the border;
on the other hand, if too much apodization is used, then one is effectively using a smaller patch of sky thus
decreasing the signal to noise.
We will use an improved version of this argument to find an optimal choice of $W$ in section \ref{sec:conjecture}. 
Here we just want to mention that when dealing with $\chi_B$, because of its very blue nature, we are forced to confront 
the issue of apodization straight on.

At this point the reader may wonder if using $\chi_B$ is such a good idea as one has to take derivatives of the data, deal with extremely blue spectra, etc. 
In the next section we will show that our technique is equivalent to the ``pure'' pseudo-$C_\ell$ technique. 
Thus all the intuition we gain focusing on $\chi_B$ will be directly translated to that other method.
The pure pseudo-$C_\ell$ formalism also clarifies the issue of statistical weight; we will see that, even though the apodized window may
suggest that the edges of the survey are downweighted, the statistical weight is in fact roughly uniformly distributed for the optimal choice
of apodization.

\section{Relation to the pure pseudo-$C_\ell$'s}
\label{sec:purecl}

We have now seen that the problem of estimating the $B$-mode power spectrum 
(without contamination from $E$-modes) can be reduced to estimating the power spectrum of a scalar 
field, by passing to the curl $\chi_B$.
In principle, any method for power spectrum estimation can be applied to $\chi_B$, to obtain
$B$-mode estimators which do not suffer from $E\rightarrow B$ mixing.
We will consider a special case which will be the focus of the rest of the paper:
estimating the power spectrum of $\chi_B$ using pseudo-$C_\ell$ estimators.

Let us recall how pseudo-$C_\ell$ power spectrum estimators are defined for the scalar field $\chi_B$.
First, one computes multipoles of the weighted field (or ``pseudo multipoles''):
\be
\ta^\chi_{\ell m} \eqdef \frac{1}{\sqrt{(\ell-1)\ell(\ell+1)(\ell+2)}} 
           \int d^2x\, \chi_B(x) W(x) Y_{\ell m}^*(x)  \label{eq:talm}
\ee
where $W(x)$ is a heuristically chosen weight function.
The issue of choosing $W(x)$ will be studied in detail in subsequent sections.
In~(\ref{eq:talm}), we have included the prefactor $1/\sqrt{(\ell-1)\ell(\ell+1)(\ell+2)}$ so that the
normalization will match the $B$-mode power spectrum $C_\ell^{BB}$.

We briefly describe the construction of unbiased bandpower estimators from the pseudo multipoles~(\ref{eq:talm}); 
for more details see \cite{Wandelt:1998qd,Hivon:2001jp}.
First we sum the multipoles in bins, obtaining pseudo bandpowers
\be
\tC_\alpha \eqdef \sum_{\ell\in\,b} \frac{\ell(\ell+1)}{2\pi} 
       \sum_{m=-\ell}^\ell \ta_{\ell m}^{\chi *} \ta_{\ell m}^{\chi}  \label{eq:pseudoC}
\ee
where the index $\alpha$ runs over $\ell$ bands.
Then we define unbiased estimators by
\be
\hC_\alpha = K_{\alpha\alpha'}^{-1} (\tC_{\alpha'} - N_{\alpha'})   \label{eq:transfer}
\ee
where $N_\alpha$ is the (additive) noise bias of each $\tC_\alpha$, and $K_{\alpha\alpha'}$ is the
transfer matrix between pseudo bandpowers~(\ref{eq:pseudoC}) and signal bandpowers $\Delta_{\alpha'}$.
(More precisely, $N$ and $K$ are defined by $\langle \tC_\alpha \rangle = K_{\alpha\alpha'} \Delta_{\alpha'} + N_\alpha$.)

The preceding construction defines a power spectrum estimator for polarization which eliminates
$E\rightarrow B$ mixing in the strongest possible sense: the estimated $B$-mode power acquires no
contribution from $E$-modes.
In \cite{Smith:2005gi}, ``pure pseudo-$C_\ell$ estimators'' for polarization, which also eliminate
$E\rightarrow B$ mixing in this strong sense, were defined.
We now prove that the two estimators are mathematically equivalent.

Substituting the definition~(\ref{eq:chiB}) of $\chi_B$ into the definition of the multipole~(\ref{eq:talm}) 
and integrating by parts, one obtains:
\ba
\ta^\chi_{\ell m} &=& \frac{i}{2}
\int d^2x\, (Q(x)+iU(x))
 \lo\lo \frac{W(x) Y_{\ell m}^*(x)}{\sqrt{(\ell-1)\ell(\ell+1)(\ell+2)}} \, + \cc    \nn  \\
&=& \frac{i}{2} \int d^2x\, (Q(x)+iU(x)) \Big[ W(x) (\Yp2^*_{\ell m}(x)) 
              + \frac{2}{\sqrt{(\ell-1)(\ell+2)}} W_1^*(x) ({}_1Y_{\ell m}^*(\bx))     \nn  \\
 && \qfive   + \frac{1}{\sqrt{(\ell-1)\ell(\ell+1)(\ell+2)}} W_2^*(x) Y^*_{\ell m}(\bx) \Big]  \, + \cc \label{eq:talm2}
\ea
where we have defined the spin-1 and spin-2 quantities
\be
W_1(x) = \ra W(x)  \qquad  W_2(x) = \ra\ra W(x)\,.  \label{eq:spin12}
\ee
In the form~(\ref{eq:talm2}), it is seen that $\ta^\chi_{\ell m}$ agrees with the pure multipole
defined in \cite{Smith:2005gi}, after changing from tensor notation to the spin-$s$ notation used here.
This shows that the pseudo multipoles in the two constructions are equal;
since the rest of the pseudo-$C_\ell$ construction 
(Eqs.~(\ref{eq:pseudoC}),~(\ref{eq:transfer})) is also the same in both cases, this completes the proof that
the $\chi_B$ estimator and the pure pseudo-$C_\ell$ estimator are equal.

Since the estimators are mathematically equivalent, it is a matter of preference which to use,
and we will switch between the two formalisms in the rest of this paper depending on which
is more convenient.
Using the $\chi_B$ estimator has the advantage that pure $B$-mode power spectrum estimation is equivalent
to power spectrum estimation of a scalar field with a very blue spectrum.
As we will see shortly, this will allow us to formulate a unified procedure for optimizing the weight function $W(x)$, 
which applies to both pure and ordinary pseudo-$C_\ell$ estimators.

In the pure pseudo-$C_\ell$ formalism, the terms have been organized (Eq.~(\ref{eq:talm2})) so that the 
derivatives act only the weight function, rather than the spherical harmonic or the noisy polarization map.  
Provided that the weight function varies slowly compared to the pixel scale, this can make it easier to
take derivatives numerically; one only has to compute $W_1(x), W_2(x)$.
The first term in Eq.~(\ref{eq:talm2}) is the usual pseudo-$C_\ell$ estimator for $B$-modes; the remaining two terms
are interpreted as higher-spin counterterms which cancel the $E\rightarrow B$ mixing.
We will see that this form of the estimator is convenient when optimizing the weight function
numerically given a noise map~(\S\ref{sec:numerical}).

We conclude this section by discussing a general property of the estimator:
the weight function must be apodized so that both $W(x)$ and its derivative vanish
at the boundary.
It is illuminating to see how this requirement arises from both ways of defining
the estimator.

For the pure pseudo-$C_\ell$ estimator, it is seen (from the mode in brackets in Eq.~(\ref{eq:talm2}),
which multiplies the Stokes parameters in a pixel) that the statistical weight of a pixel is given by 
a combination of $W(x)$ and its first two derivatives.
Therefore, if either $W(x)$ or its derivative have nonzero boundary values, the statistical weight
will contain a delta function on the boundary, which leads to a divergent estimator.
(One might try to cure this divergence by defining $W_1, W_2$ with the delta function terms omitted
from the derivatives; the estimator will then be finite, but since the relations~(\ref{eq:spin12}) do not hold
strictly, the estimator will mix $E\rightarrow B$.)
This reasoning also shows that even though $W(x)$ is apodized near the boundary, the statistical weight of pixels 
near the boundary need not be small, since the counterterms $W_1$ and $W_2$ will be largest there.
In fact, for the optimized weight functions that we will consider in subsequent sections, we have found that
the statistical weight is roughly uniformly distributed.

For the $\chi_B$ estimator, the divergence arises in a different way.
Although the estimator appears not to involve derivatives of $W(x)$, the
field $\chi_B$ has a noise power spectrum on small scales which grows as $\ell^4$.
If either $W(x)$ or its derivative is nonzero on the boundary,
then $||\tW(l)||^2$ decays as $1/\ell^4$ or slower for large $l$ (Eq.~(\ref{eq:nderivs}))
which causes the integral~(\ref{cl}) to diverge.
From this perspective, the need to apodize is understood as a consequence of small-scale noise fluctuations
in $\chi_B$ with a very blue power spectrum.

\section{Choosing the optimal weight function}
\label{sec:conjecture}

In pseudo-$C_\ell$ power spectrum estimation, an important practical issue is choosing
the pixel weight function $W(x)$ to minimize the variance of the estimator.
Although no general procedure has been proposed, several rules of thumb for optimizing $W(x)$ 
under different limiting conditions have appeared.
In the noise-dominated limit, inverse noise weighting ($W(x) = 1/\sigma^2(x)$) is
optimal; in the signal-dominated limit, uniform weighting is best possible \cite{Challinor:2004pr}.
If the power spectrum is being estimated on scales smaller than fluctuations in the noise level, the FKP
ansatz \cite{Feldman:1993ky} has become the industry standard for galaxy surveys.

The preceding rules of thumb implicitly assume that the signal + noise power spectrum is white on small scales.
However, we have seen that when choosing the weight function for pure $B$-mode estimation, one includes an
extra factor $\ell^4$ in the power spectrum.  This qualitatively affects the optimization and invalidates
the rules of thumb.  For example, $W(x)$ and its first derivative must vanish on survey boundaries, in contrast
to the white noise case.

In this section, we will describe a general ansatz for the optimal weight function, which makes
sense for an arbitrary $\npix$-by-$\npix$ matrix $\bC$ representing the total covariance (signal plus noise).
In particular, it can be applied to a noise power spectrum which grows as any power of $\ell$, and so provides a uniform
framework for both pure and ordinary pseudo-$C_\ell$ power spectrum estimation.
We will see that our ansatz reproduces the above rules of thumb in the appropriate limits,
and correctly apodizes $W(x)$ if the power spectrum is blue on small scales.

First we introduce some notation.
The $\npix$-by-$\npix$ signal covariance in the bandpower we are estimating
will be denoted $\bCa$, where $\alpha$ is an index ranging over $\ell$ bands, and we normalize it
so that $\bCa_{ii}=1$.

If $M_1$ and $M_2$ are two $\npix$-by-$\npix$ matrices, then $[M_1 * M_2]$ denotes the $\npix$-by-$\npix$
matrix obtained by {\em elementwise} multiplication:
\be
[M_1 * M_2]_{ij} = (M_1)_{ij} (M_2)_{ij}\,.  \label{eq:star}
\ee
The multiplication is always performed in a basis where each row or column corresponds to one pixel;
otherwise the operation defined by (\ref{eq:star}) would be basis-dependent.

Now we can state the ansatz for optimizing the pseudo-$C_\ell$ weight function
which will be the main result of this paper:
the optimal weight function $\Wopt$ (thought of as a length-$\npix$ vector)
is given by
\be
\Wopt = [\bCa * \bC]^{-1} \one            \label{eq:Wopt}
\ee
where $\one$ deontes the length-$\npix$ vector consisting of all 1's.

It may seem strange that the ansatz~(\ref{eq:Wopt}) includes the * operation~(\ref{eq:star}) which is
specific to the pixel-space basis.  Loosely speaking, in pseudo-$C_\ell$ power spectrum estimation, we
are constrained to use a filter (multiplying by the mask) which is diagonal in pixel space, and so the
pixel basis is given special significance.
Before giving the derivation of the ansatz, we note some general properties.
Without using the * operation, it seems to be impossible to achieve all of the following at once:
\begin{enumerate}
\item If the noise is uncorrelated between pixels, and the noise dominates, then $\Wopt$ corresponds 
to inverse noise weighting.  
This follows from Eq.~(\ref{eq:Wopt}) upon noting that $\bC$ is diagonal in the pixel basis, 
say $\bC_{ij}=\sigma_i^2\delta_{ij}$, 
so that $(\bCa * \bC)=\bC$ and $(\Wopt)_i = \sigma_i^{-2}$.
Note that in this case, pseudo-$C_\ell$ power spectrum estimation is optimal, in the sense that the Cramer-Rao
inequality is saturated.
\item If the signal + noise power spectrum grows as a power law $\ell^\lambda$ on small scales, and $\lambda > 0$,
then $\Wopt$ is apodized near the boundary.
The number of derivatives which vanish at the boundary is given by $(\lambda/2 - 1)$, in accord with
the discussion at the end of \S\ref{ssec:apodization}.
As the wavenumber $\ell_0$ where one is estimating the power spectrum increases, the apodization length 
decreases.
These statements will be proved in \S\ref{sec:MZ}.
\item Eq.~(\ref{eq:Wopt}) reduces to the FKP prescription in the regime where FKP applies.  This is shown in Appendix~\ref{sec:FKP}.
\end{enumerate}
The matrix form of the ansatz~(\ref{eq:Wopt}) may suggest that $\npix$-by-$\npix$ matrix operations are needed to compute
the optimal weight function, which would be problematic since such operations are computationally prohibitive in situations
where pseudo-$C_\ell$ estimators are used.  (If they were feasible, then one could do a maximum likelihood analysis instead.)
However, in Appendix \ref{app:cg}, we will see that a conjugate gradient approach may be used to compute the RHS of Eq.~(\ref{eq:Wopt})
even in surveys where the pixel count is large.

\subsection{Derivation}
\label{sec:KMS}

Now we give a heuristic derivation of the ansatz~(\ref{eq:Wopt}).
The idea is to find the weight function $W$ such that the pseudo-$C_\ell$ estimator (defined in Eq.~(\ref{eq:pseudoC}))
\be
\tC_\alpha[\bd] = \sum_{ij} \bd_i W_i \bCa_{ij} W_j \bd_j  \label{eq:KMS0}
\ee
is as close as possible to the optimal estimator
\be
\tO_\alpha[\bd] \eqdef \sum_{ij} \bd_i (\bC^{-1} \bCa \bC^{-1})_{ij} \bd_j\,.    \label{eq:KMS0b}
\ee
In Eqs.~(\ref{eq:KMS0}),~(\ref{eq:KMS0b}) we have denoted the data vector (a length-$\npix$ vector whose covariance matrix is $\bC$) by $\bd$.

How should ``close'' be defined?
We first note that both estimators can be written as Monte Carlo averages
\ba
\tC_\alpha[\bd] &=& \left\langle \left( \sum_i \bp_i W_i \bd_i \right)^2 \right\rangle_{\bp}    \nn   \\
\tO_\alpha[\bd] &=& \left\langle \left( \sum_{ij} \bp_i \bC^{-1}_{ij} \bd_j \right)^2  \right\rangle_{\bp}  \label{eq:KMS1}
\ea
where the notation $\langle \cdot \rangle_{\bp}$ means that an expectation value is taken over
an auxiliary field $\bp$ with covariance $\bCa$.
From the expressions~(\ref{eq:KMS1}), a natural definition of ``close'' is that the expectation value
\be
E \eqdef \left\langle 
\left( \sum_i \bp_i W_i \bd_i - \sum_{ij} \bp_i \bC^{-1}_{ij} \bd_j \right)^2
\right\rangle_{\bd,\bp}   \label{eq:KMS3}
\ee
be minimized.  Here, the expectation value is taken over both the auxiliary field $\bp$ with covariance
$\bCa$ and the data vector $\bd$ with covariance $\bC$.
We minimize $E$ by setting its derivative with respect to the weight function to zero:
\ba
0 = \frac{1}{2} \frac{\partial E}{\partial W_k} &=& \left\langle
   \bp_k \bd_k \left(\sum_i \bp_i W_i \bd_i - \sum_{ij} \bp_i \bC^{-1}_{ij} \bd_j \right)
   \right\rangle_{\bd,\bp}  \nn \\
 &=& \sum_i \bCa_{ik} \bC_{ik} W_i - \sum_{ij} \bCa_{ik} \bC^{-1}_{ij} \bC_{jk}  \nn \\
 &=& \sum_i \bCa_{ik} \bC_{ik} W_i - 1  \qquad\mbox{(no sum on $k$).}
\ea
i.e., $\Wopt = (\bC * \bCa)^{-1} \one$.  This completes the derivation of~(\ref{eq:Wopt}).

\subsection{A variational form of the ansatz}
\label{sec:MZ}

In Eq.~(\ref{eq:Wopt}), we have written the ansatz for the optimal weight function $\Wopt$ as a matrix equation.
There is an equivalent formulation as a variational principle which connects to the discussion in \S\ref{ssec:apodization} 
and will also be directly useful in later sections: $\Wopt$ is the weight function which minimizes the expectation value of
the pseudo bandpower $\tC_\alpha$
\ba
\langle \tC_\alpha \rangle &=& \left\langle \sum_{ij} \bd_i W_i \bCa_{ij} W_j \bd_j  \right\rangle  \nn  \\
                           &=& \sum_{ij} W_i W_j \bC_{ij} \bCa_{ij}  \label{eq:MZ1}
\ea
subject to the normalization condition
\be
\sum_i W_i = \mbox{const.}   \label{eq:MZ2},
\ee
This is shown by differentiating Eq.~(\ref{eq:MZ1}) with respect to $W_i$, obtaining
\be
\sum_j \bC_{ij} \bCa_{ij} W_j = 1 \qquad\mbox{(no sum on $i$)}  \label{ansatzeq}
\ee
i.e., $\Wopt = (\bC * \bCa)^{-1} \one$.  (Strictly speaking, the RHS in Eq.~(\ref{ansatzeq}) should be equal to
a Lagrange multiplier $\lambda$, but the overall normalization of the weight function is arbitrary.)

In the previous subsection we showed that $\Wopt$ is the weight function which makes the pseudo-$C_\ell$
estimator approximate the optimal estimator as closely as possible.
The variational principle gives another interpretation of $\Wopt$ which matches the
discussion from \S\ref{ssec:apodization}: $\Wopt$ is the weight function which minimizes 
the total aliasing from both signal outside the bandpower and noise.
Even if known perfectly on average, say because the power spectrum on other scales was given,
this aliasing will contribute to the noise in the estimator thus degrading the signal to noise.

The variational principle also gives a simple proof of property \#2 at the end of \S\ref{sec:conjecture},
that $(\lambda/2 - 1)$ derivatives of $\Wopt$ vanish at the boundary if the power
spectrum is a power law $\ell^\lambda$ on small scales.
From Eq.~(\ref{eq:nderivs}), if $n$ is the number of derivatives which vanish, then the contribution
to the expectation value $\langle \tC_\alpha \rangle$ from small scales (large $l$) is given by
\be
\langle \tC_\alpha \rangle \propto \int \frac{d^2l}{(2\pi)^2}\, l^{\lambda - 2n - 4}
\ee
Therefore, $n \ge (\lambda/2 - 1)$ is a necessary condition for the integral to converge,
and so $\Wopt$ must satisfy this condition, since it is chosen to minimize the expectation
value.

\section{Some analytic solutions}
\label{sec:analytic}

To build intuition for the ansatz~(\ref{eq:Wopt}), we give some analytic solutions
in the case where the survey region is a circle of radius $R$ and the covariance $\bC$ is
given by a power-law power spectrum $\ell^\lambda$, where $\lambda=2$ or 4.
We assume uniform noise throughout the survey region, so that the covariance is completely
described by a power spectrum, but statistical isotropy is broken by the survey boundary.
More realistic surveys and power spectra will be considered in subsequent sections.
We use the flat sky approximation throughout.

\subsection{Case 1: $\ell^2$ power spectrum}

Let us first consider the case where the power spectrum is proportional to $\ell^2$,
so that the operator $\bC$ is equal to $(-\nabla^2)$.  Computing the operator $(\bC * \bC_\alpha)$
involves some subtleties, so we show the details explicitly.  In the position space basis, the
matrix elements of $\bC$ and $\bC_\alpha$ are given by
\ba
(\bC)_{xy} &=& -\nabla^2 \delta^2(x-y)  \nn  \\
(\bC_\alpha)_{xy} &=& J_0(\ell_0|x-y|)\,.
\ea
where $\ell_0$ denotes the wavenumber where we are estimating the power spectrum.
From the definition~(\ref{eq:star}), the matrix elements of $(\bC * \bC_\alpha)$ are
\ba
(\bC * \bC_\alpha)_{xy} &=& \Big[-\nabla^2 \delta^2(x-y) \Big] J_0(\ell_0|x-y|)  \nn \\
                        &=& -\nabla^2\delta^2(x-y) + \ell_0^2\delta^2(x-y)   \label{eq:l2op} \\
\ea
where we have used $J_0(0)=1$, $J'(0)=0$, $J_0''(0)=-1/2$.
This calculation shows that the operator $(\bC * \bC_\alpha)$ is equal to $(-\nabla^2 + \ell_0^2)$.

\begin{figure}
\centerline{\epsfxsize=7.0truein\epsffile[18 480 592 718]{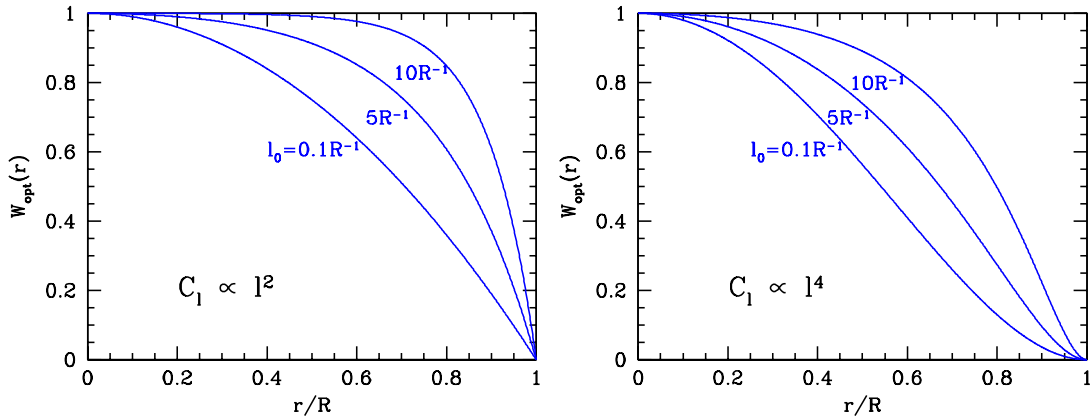}}
\caption{Optimal window functions $W_{\rm opt}(r)$, given by Eqs.~(\ref{eq:l2sol}),~(\ref{eq:l4sol})
for a signal+noise power spectrum proportional to $l^2$ (left panel) or $l^4$ (right panel).}
\label{fig:analytic}
\end{figure}

Combining Eqs.~(\ref{eq:Wopt}),~(\ref{eq:l2op}), the optimal weight function satisfies the differential equation
\be
(-\nabla^2 + \ell_0^2) W_{\rm opt}(x) = \mbox{const.}    \label{eq:l2de}
\ee
with Dirichlet boundary conditions (as proved in \S\ref{sec:MZ}).  The solution to~(\ref{eq:l2de}) is given by
\be
W_{\rm opt}(r) = 1 - \frac{I_0(\ell_0r)}{I_0(\ell_0R)}\,.   \label{eq:l2sol}
\ee
The solutions to~(\ref{eq:l2sol}) are shown in the left panel of Fig.~\ref{fig:analytic}.
For the $\lambda=2$ power spectrum, the weight functions are apodized near the boundaries with an apodization
length which decreases with increasing $\ell_0$.
This is in contrast to the white noise case where the weight function would be uniform, independently of $\ell_0$,
if the noise is homogeneous throughout the survey.

\subsection{Case 2: $l^4$ power spectrum}

We next consider the case of a signal + noise power spectrum which is proportional to $\ell^4$.
This case corresponds to pure $B$-mode estimation in the noise-dominated limit.

A calculation similar to~(\ref{eq:l2op}) shows that the differential equation for the weight function is
\be
(-\nabla^4 + 4 \ell_0^2 \nabla^2 - \ell_0^4) W = \mbox{const.}  \label{eq:l4de}
\ee
with Dirichlet + Neumann boundary conditions.
The solution to~(\ref{eq:l4de}) is
\be
W_{\rm opt}(r) = 1 - \frac{\ell_+I_0(\ell_-r)I_1(\ell_+R) - \ell_-I_0(\ell_+r)I_1(\ell_-R)}
                          {\ell_+I_0(\ell_-R)I_1(\ell_+R) - \ell_-I_0(\ell_+R)I_1(\ell_-R)}
\qquad\mbox{where $\ell_\pm \eqdef (2 \pm \sqrt{3})^{1/2} \ell_0$.}   \label{eq:l4sol}
\ee
This solution is shown in the right panel of Fig.~\ref{fig:analytic}.
Relative to the $\lambda=2$ case (left panel), the apodization length is larger, and the apodization
is such that both $W(x)$ and its derivative vanish on the boundary.

\section{Numerical solutions}
\label{sec:numerical}

Let us summarize our results so far.
We have shown that $E\rightarrow B$ mixing in pseudo-$C_\ell$ power spectrum estimation can be eliminated
by taking the curl of the map, then estimating the power spectrum of the resulting scalar field.
Because the curl operation results in a noise power spectrum which grows as $C_\ell \propto \ell^4$,
the weight function must be smoothly apodized near boundaries in order 
to control contamination from small-scale power.
This technique is mathematically equivalent to the pure pseudo-$C_\ell$ formalism from \cite{Smith:2005gi},
in which ``counterterms'' involving spin-1 and spin-2 weights
are added to the ordinary pseudo-$C_\ell$ estimator to cancel $E\rightarrow B$ mixing.
We have proposed a general ansatz~(\ref{eq:Wopt}) for optimizing pseudo-$C_\ell$ weight functions in the
presence of arbitrary signal and noise covariance, and constructed analytic solutions in special cases.

For practical application, one needs a method for solving the ansatz numerically, giving a specification
of the noise.
The most general noise model we will consider is uncorrelated between pixels, isotropic in 
each pixel, but with an arbitrary pixel-dependent amplitude:
\be
\left( \begin{array}{cc}
\langle Q(x) Q(x') \rangle  &  \langle Q(x) U(x') \rangle  \\
\langle U(x) Q(x') \rangle  &  \langle U(x) U(x') \rangle
\end{array} \right)
=
\left( \begin{array}{cc}
\sigma^2(x) & 0  \\
0 & \sigma^2(x)
\end{array} \right) \delta_{xx'}    \label{eq:uncorr}
\ee
For this noise model, we have implemented a ``black-box'' procedure which starts from the noise RMS $\sigma(x)$
in each pixel and outputs $E$-mode and $B$-mode weight functions in each $\ell$ band.
The procedure uses the variational principle from \S\ref{sec:MZ}, finding weight functions which minimize
the quantity $\langle \tC_\alpha \rangle$.
In this and the next section, we will describe qualitative features of the solutions and study estimator 
performance, deferring implementational details of the method to Appendix~\ref{app:cg}.

We work in the pure pseudo-$C_\ell$ formalism, with one minor modification.
As we have described it in \S\ref{sec:purecl}, the spin-1 and spin-2 weights are always given in
terms of the spin-0 piece by:
\be
W_1(x) = \ra W(x)  \qquad\qquad W_2(x) = \ra\ra W(x) \, ,  \label{eq:constraints}
\ee
In our numerical optimization procedure, we do not impose Eqs.~(\ref{eq:constraints}) as constraints; 
each $B$-mode ``weight function'' actually consists of three independent pieces:
a spin-0 (scalar) function $W_0(x)$, a spin-1 (vector) field $W_1(x)$, and a spin-2 (tensor) field $W_2(x)$.
Note that for EE power spectrum estimation, we do not include higher-spin counterterms, and so each $E$-mode weight
function is simply a scalar $W(x)$.

\begin{figure}
\centerline{\epsfxsize=6.2truein\epsffile{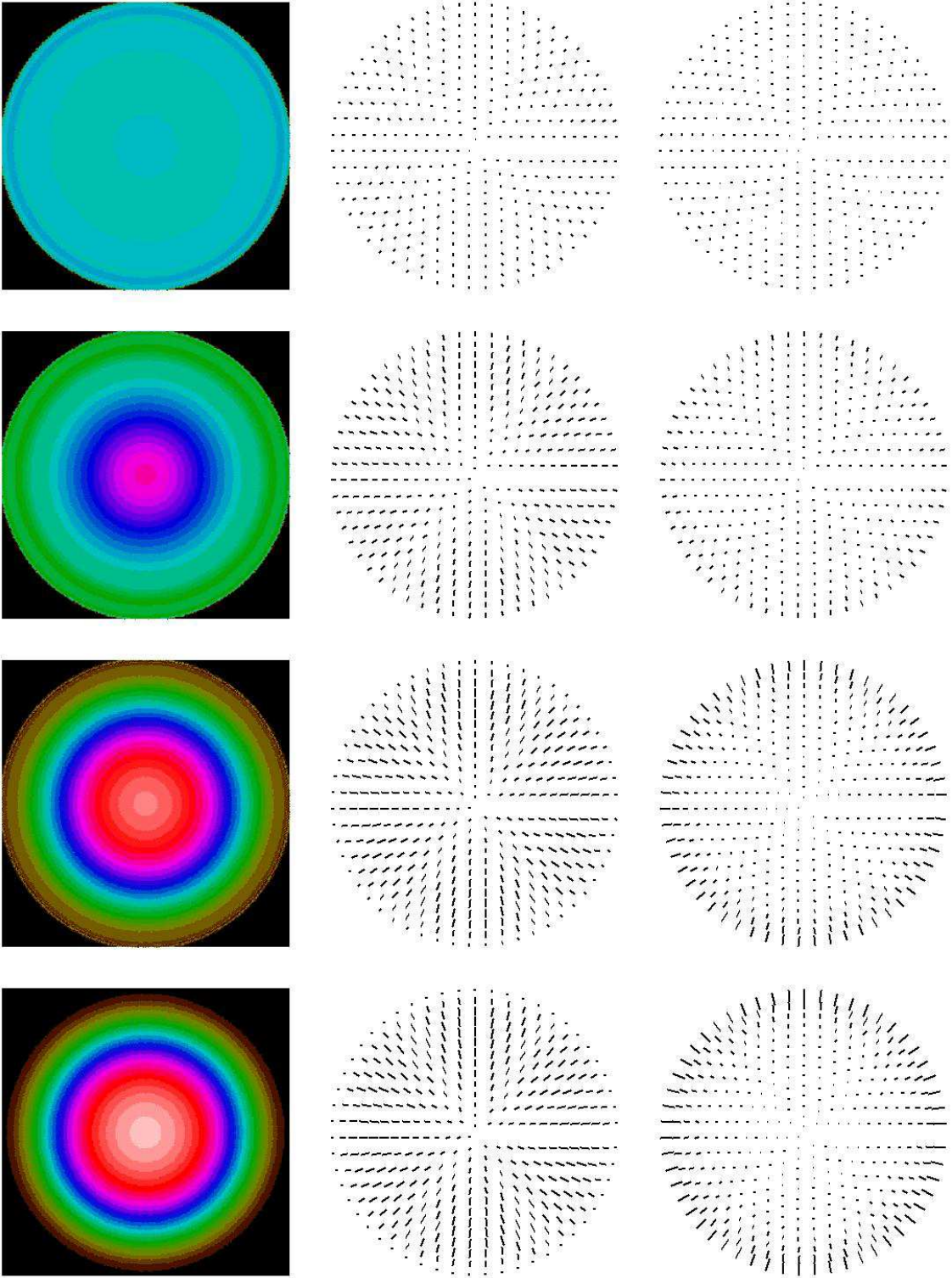}}
\caption{$B$-mode weight functions for varying noise levels, for $\ell=20$ and a $10^\circ$ spherical cap with homogeneous noise.
As descrbed in the text, each weight function consists of spin-0, spin-1, and spin-2 pieces, which we show in the
left, center, and right columns.  The top three rows show the $B$-mode weight functions produced by our
optimization procedure for noise levels 100, 20, and 5 $\mu$K-arcmin (from top to bottom).
In the bottommost row we show the analytic solution given by Eq.~(\ref{eq:l4sol}).}
\label{fig:noiselevels}
\end{figure}

In order to differentiate between different flavors of estimators, we will use the following
terminology in the rest of the paper.
If both counterterms are absent (i.e., $W_1 = W_2 = 0$), we will refer to the $B$-mode estimator as ``ordinary pseudo-$C_\ell$''.
We will reserve the term ``pure pseudo-$C_\ell$'' for the situation where the relations~(\ref{eq:constraints}) hold exactly,
e.g. because $W(x)$ is of known analytical form and $W_1, W_2$ are obtained by simply evaluating derivatives in each pixel.
In this case the $B$-mode estimator will have zero $E\rightarrow B$ leakage (except perhaps from aliasing artifacts in a finite
pixelization).
Finally, in a case where the relations~(\ref{eq:constraints}) hold approximately, we will refer to the $B$-mode estimator as
``pseudo-$C_\ell$ with counterterms''.
This will be the situation for the weight functions produced by our numerical optimization procedure; since the 
relations~(\ref{eq:constraints}) are not imposed as constraints, they do not hold strictly, and so there will be some
nonzero mixing in the estimator.
However, we will see shortly that the relations will hold approximately, and the level of $E\rightarrow B$ mixing for
optimized weight functions is small in practice.
This is because reducing the mixing helps minimize the quantity $\langle \tC_\alpha \rangle$, so our optimization procedure
prefers weight functions which satisfy Eqs.~(\ref{eq:constraints}) to a good approximation.
In fact, this is the main reason why we have found it convenient to work in the counterterm formalism,
rather than passing to the curl $\chi_B$ (which we showed was equivalent in \S\ref{sec:purecl}): 
the variational optimization procedure sidesteps implementational issues associated with taking derivatives in an irregular 
spherical pixelization such as Healpix.

In Fig.~\ref{fig:noiselevels}, we show the result of our optimization procedure for $\ell=20$, a spherical cap shaped survey
with radius $10^\circ$, and three choices of noise level: 100, 20, and 5 $\mu$K-arcmin.
For the largest noise level, both higher-spin counterterms are small and the weighting is roughly uniform.
In this regime, our optimization procedure reduces to ordinary pseudo-$C_\ell$ power spectrum estimation
with tophat weighting.
For the smallest noise level, the weight function is apodized, both counterterms are present, and all components 
of the weight function have nearly converged to the analytic solution~(\ref{eq:l4sol}), which represents
the limit $EE \gg BB$.
Here our procedure reduces to pure pseudo-$C_\ell$ with weighting that can be understood by solving
the differential equation~(\ref{eq:l4de}).
Our optimization procedure therefore smoothly interpolates between ordinary and pure pseudo-$C_\ell$
as the noise level is improved, ``turning on'' the higher-spin counterterms as they are needed to reduce
$E\rightarrow B$ mixing below the noise floor.  (We will illustrate this in a different way in the next section.)

\begin{figure}
\centerline{\epsfxsize=7.0truein\epsffile[18 330 592 718]{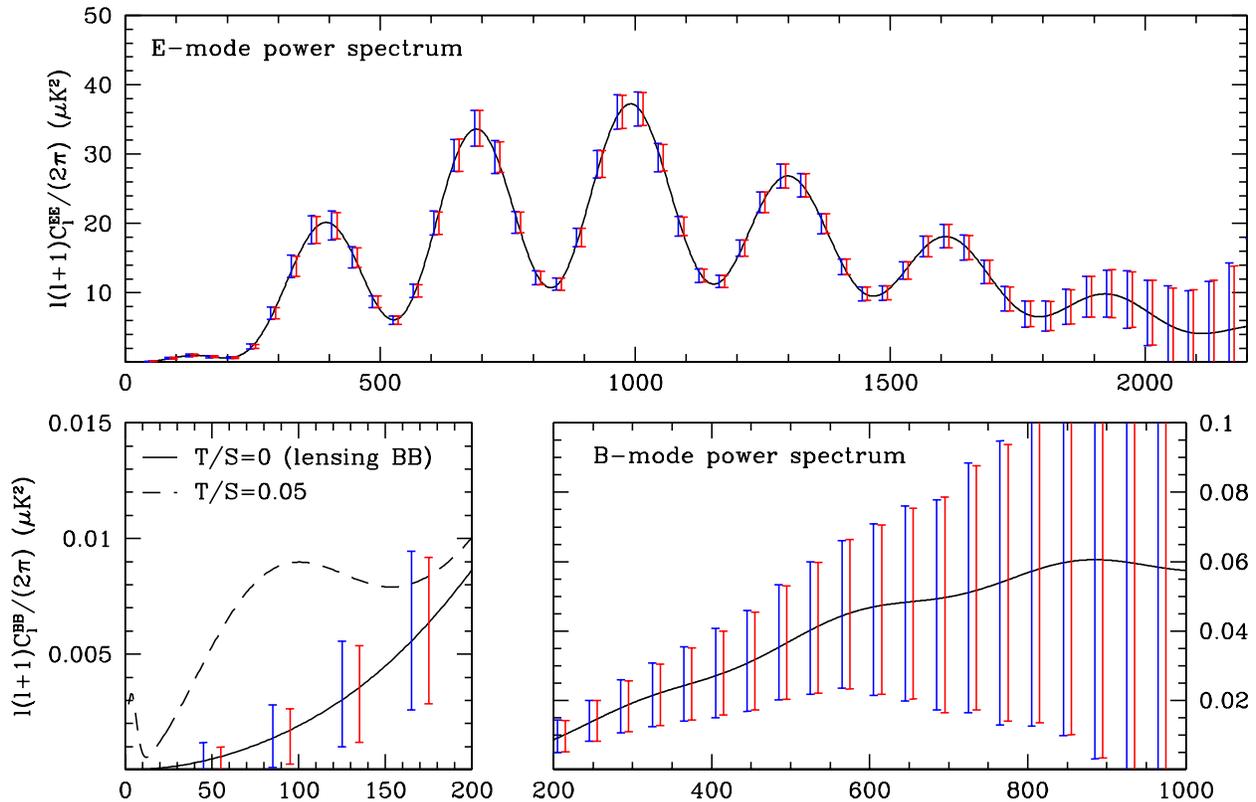}}
\caption{Power spectrum errors for the $10^\circ$ homogeneous mock survey described in \S\ref{sec:numerical},
using pseudo-$C_\ell$ power spectrum estimation with counterterms (left/blue) or optimal estimators (right/red).
For the $B$-mode power spectrum, we compute power spectrum errors in a fiducial model with $T/S=0$
but also show a spectrum with $T/S=0.05$ for comparison.}
\label{fig:stcerr}
\end{figure}

Next we consider power spectrum errors from our estimators for a spherical cap shaped mock survey with
radius $10^\circ$, homogeneous noise level 5.75 $\mu$K-arcmin and Gaussian beam $\theta_{\rm FWHM} = 8$ arcmin.
These values have been chosen to roughly model the fiducial realistic survey
which will be treated in more detail in the next section.

We have used an azimuthally symmetric mock survey so that optimal, or maximum likelihood, power
spectrum estimation will be computationally feasible, even though the pixel count is large enough
that this would normally be impossible.
The computational speedup is obtained because both signal and noise covariance matrices are diagonal
in $m$; for details see Appendix~F of \cite{Smith:2005gi}.
This allows us to compare our estimators to optimal in a baseline survey which approximates
our fiducial realistic experiment (\S\ref{sec:ebex}) within the constraint of azimuthal symmetry.

In Fig.~\ref{fig:stcerr}, we show power spectrum errors for the mock survey using both pseudo-$C_\ell$ with weight
functions produced by our optimization procedure, and optimal power spectrum estimators.
It is seen that our method is slightly suboptimal at very low $\ell$ but rapidly becomes optimal.
For example, the $B$-mode bandpower RMS is 72\% optimal in the lowest band, 88\% optimal in the second-lowest,
and 91\% optimal in the third lowest.
We emphasize that this level of performance for $B$-modes is much better than could be obtained at this noise level using
ordinary pseudo-$C_\ell$ estimators \cite{Challinor:2004pr}, and that we have obtained it using a completely 
automated procedure for generating weight functions from the noise distribution.

\section{A realistic example}
\label{sec:ebex}

\begin{figure}
\centerline{\epsfxsize=4.5truein\epsffile{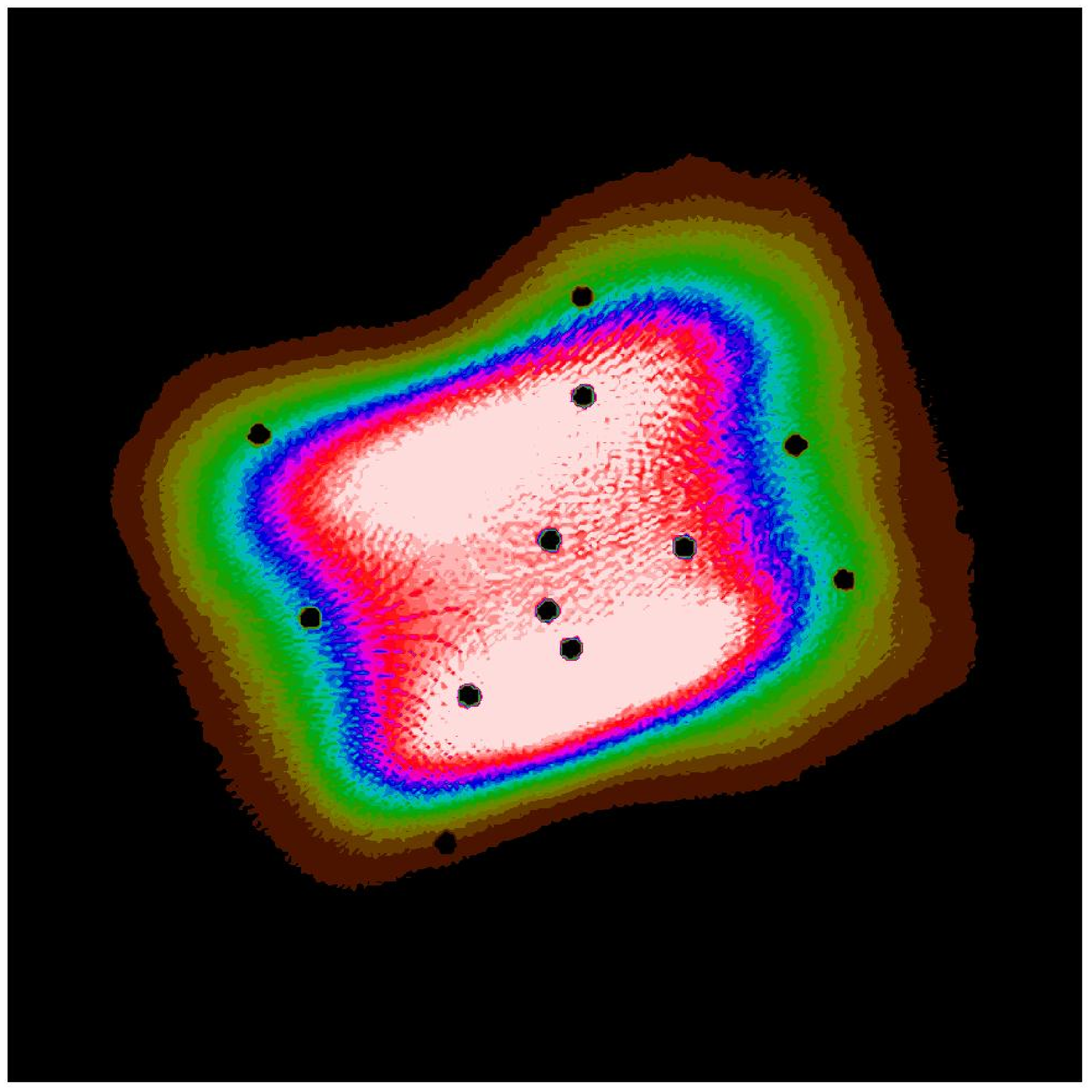}}
\centerline{\epsfxsize=4.5truein\epsffile{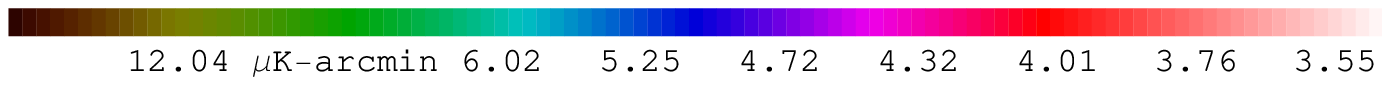}}
\caption{Survey region, noise distribution, and point source mask for our fiducial realistic experiment.  The side length of the bounding square is 25$^\circ$.  The map is based on preliminary simulations of the upcoming balloon experiment EBEX.}
\label{fig:ebex_hitcount}
\end{figure}

In the preceding section we considered a homogeneous, azimuthally symmetric mock survey whose
noise distribution was much simpler than a real CMB experiment.
We conclude this paper by considering a more realistic example. We will use noise maps made from preliminary simulations of the EBEX experiment \cite{sam}. The noise map is shown in Fig.~\ref{fig:ebex_hitcount}. The simulation used the scan strategy proposed by EBEX and a realistic focal plane configuration to 
estimate the number of hits for every 1 arcminute square pixel after a 14 day long duration 
balloon flight. 
The details of the scan, focal plane configuration, number of detectors and ultimate sensitivity  are not of particular interest here 
and are probably subject to change before the flight takes place. 
However the simulations give a good illustration of the level of complexity of the noise maps 
and sky coverage for upcoming ground and balloon based experiments. Thus they serve as a nice test bed for our methods. 
    
Since the map is not azimuthally symmetric, computing optimal power spectrum errors is
computationally prohibitive and we will not be able to compare our method to optimal.
Instead we will compare pseudo-$C_\ell$ power spectrum errors for the fiducial realistic experiment
to those obtained for the azimuthally symmetric mock survey in the preceding section,
which roughly approximates the sky coverage and noise distribution.

In Fig.~\ref{fig:ebex_hitcount}, we have generated randomly-positioned point sources 
with average number density 0.04 deg$^{-2}$, and masked each point source to radius 17 arcmin, corresponding
to the $5\sigma$ level of the beam (assumed Gaussian with $\theta_{\rm FWHM}=8$ arcmin).
With these parameters, 1\% of the survey area is excluded by the point source mask. 
The area that will need to be cut out to sufficiently mask point sources to measure $B$-mode polarization is at this point rather uncertain. 
We have based our choice of 1\% on point source simulations conducted at SISSA for the Planck satellite reference sky \cite{carlo}. 
Our choice should thus be seen as an illustrative example to help us understand how our method would deal with the resulting holes. 
We will analyze the map with and without the mask, to isolate the impact of point sources
when measuring $B$-modes.

\begin{figure}
\centerline{\epsfxsize=7.0truein\epsffile{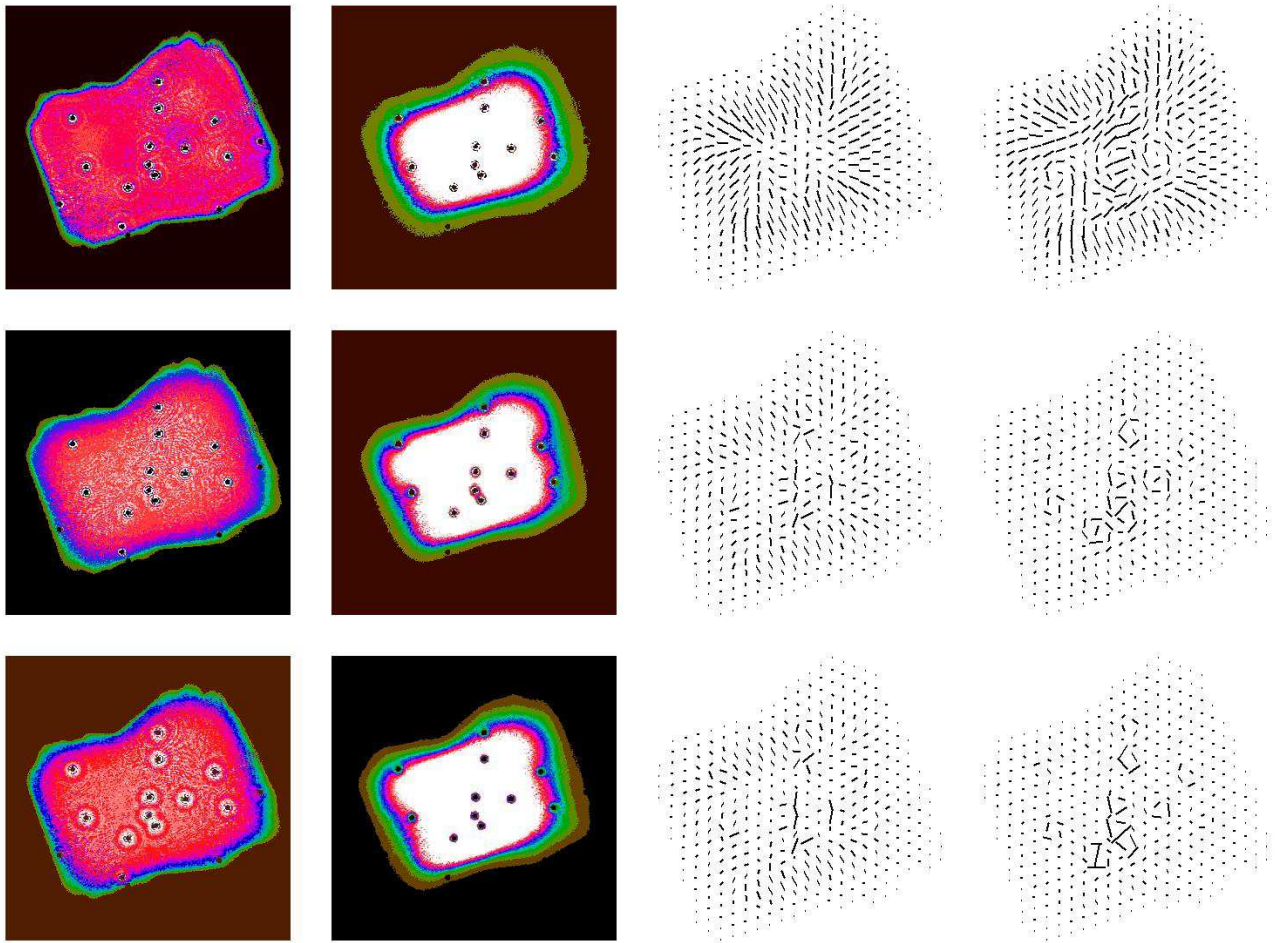}}
\caption{Weight functions for the fiducial realistic experiment.  Each of the three rows corresponds to
a different bandpower, with multipole ranges $(\ellmin,\ellmax)$ given from top to bottom
by (30,70), (190,230) and (510,550).  Within each row from left to right,
the $E$-mode weight function, the spin-0 piece of the $B$-mode weight function,
and the spin-1 and spin-2 pieces of the $B$-mode weight function are shown.}
\label{fig:ebex_weights}
\end{figure}

\begin{figure}
\centerline{\epsfxsize=6.5truein\epsffile{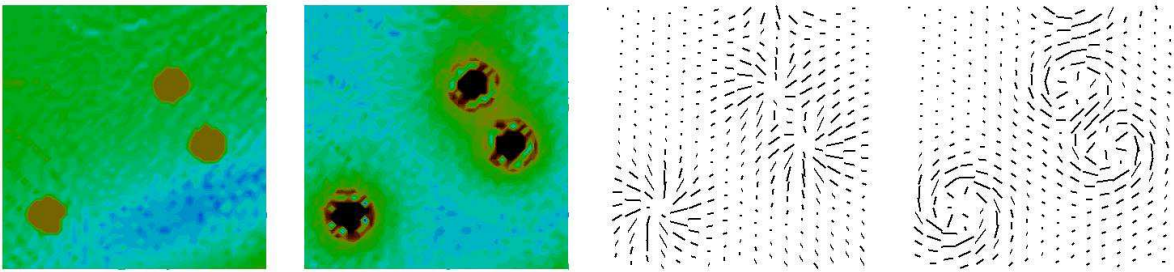}}
\caption{A zoomed view of the $(\ellmin,\ellmax) = (190,230)$ $B$-mode weight function near point sources.
From left to right, the panels show the noise map, the scalar piece of the weight function, and the two
higher-spin counterterms.}
\label{fig:ebex_zoom}
\end{figure}

For analyzing a very inhomogeneous map such as this, with noise fluctuations on large and
small scales, an automated procedure for optimizing weight functions is a practical necessity.
Using the optimization procedure described in \S\ref{sec:numerical}, we generated weight functions
for both $E$-mode and $B$-mode power spectrum estimation; some representative weight functions
are shown in Fig.~\ref{fig:ebex_weights}.

Considering $E$-mode weight functions first, our method generates weight functions which are
more inhomogeneous than those given by the FKP ansatz (Appendix~\ref{sec:FKP}) but similar in performance.
There is a slight improvement at low $\ell$; e.g. for the lowest $E$-mode bandpower, the
the bandpower RMS given by our estimator is $\sim 10\%$ better than FKP.
Note that in \S\ref{sec:numerical}, we did not discuss $E$-mode weight functions, since they are very
simple for the homogeneous mock survey.

Turning now to $B$-mode weight functions, one qualitative feature is that their statistical
weight is concentrated near the center of the survey (compared to the $E$-mode weight functions).
This is for signal-to-noise reasons, since the $B$-mode signal is weak and the center of the
survey is least noisy.

Another feature of the $B$-mode weight functions is that the higher-spin counterterms
are relatively smooth across the map, even though the scalar piece of the weight function
has structure on small scales matching similar structure in the noise.
Instead, our numerical optimization procedure prefers to associate counterterms with
``large scale'' features such as survey boundaries or point sources, but not with small-scale
features in the noise.
From Fig.~\ref{fig:ebex_weights}, one sees that the counterterms are sourced mainly by boundaries at low $\ell$ 
and by point sources at intermediate to high $\ell$.
A zoomed-in view of the $B$-mode weight function near point sources is shown in Fig.~\ref{fig:ebex_zoom}.
The behavior of the weight function near the ``internal'' boundaries of point sources is
similar to the behavior near external survey boundaries seen in Fig.~\ref{fig:noiselevels}:
the scalar piece of the weight function is apodized, and the counterterms
are large near the boundary.

\begin{figure}
\centerline{\epsfxsize=7.0truein\epsffile[18 460 592 710]{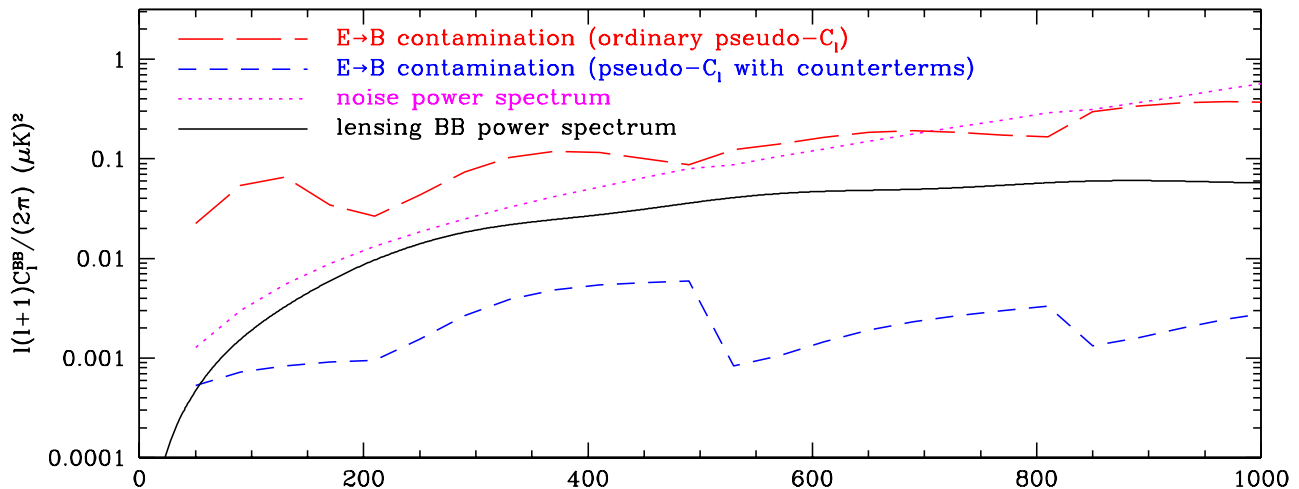}}
\caption{Contribution of aliased $E$-modes to the pseudo power spectrum $\tC^{BB}_\ell$ for the fiducial
realistic experiment with point source mask, with signal and noise power spectra shown for comparison.
Using counterterms from our numerical optimization procedure, the level of $E\rightarrow B$ mixing
is well below the noise floor, in contrast to ordinary pseudo-$C_\ell$.}
\label{fig:sbias}
\end{figure}

One consequence of the counterterms being smooth, even though the scalar piece of the $B$-mode weight functions
have small-scale structure, is that the relations~(\ref{eq:constraints}) are not strictly satisfied.
Therefore, the estimator is not completely pure; there is some nonzero level of $E\rightarrow B$ mixing.
This is quantified in Fig.~\ref{fig:sbias}, where we show the mean contribution of aliased $E$-modes to the pseudo
power spectrum $\tC_\ell$.  In the pseudo-$C_\ell$ construction, this contribution is always removed in the mean by the
debiasing step, but acts as a source of extra variance.  It is seen that our optimization procedure produces counterterms
which reduce $E\rightarrow B$ mixing well below the level of the noise, even though strictly speaking, the mixing
is nonzero, i.e. the estimator is not pure.
(In Fig.~\ref{fig:sbias}, some ``stair-step'' features can be seen which arise because we do not generate a separate set of
weight functions in every band, e.g. we use the same weight function between $\ell=510$ and $\ell=830$ to save CPU time.)

\begin{figure}
\centerline{\epsfxsize=7.0truein\epsffile[18 330 592 718]{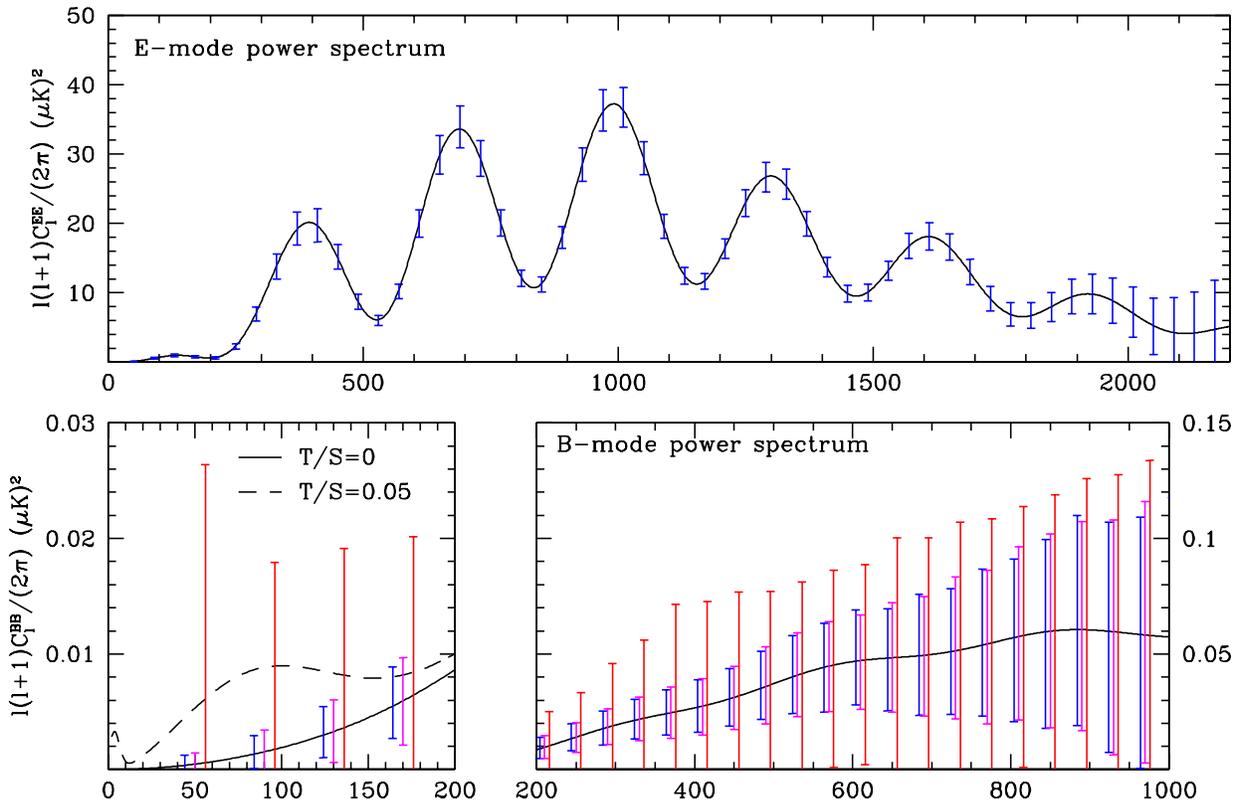}}
\caption{Comparison of three sets of pseudo-$C_\ell$ power spectrum errors for the fiducial realistic experiment.
Weight functions were optimized independently in each of the three cases.
{\em Left/blue:} Pseudo-$C_\ell$ with counterterms, without the point source mask.
{\em Center/magenta:} Pseudo-$C_\ell$ with counterterms, with the mask.
{\em Right/red:} Ordinary pseudo-$C_\ell$, with the mask.
For $E$-mode power spectrum errors, the three cases are so similar that only one set of error bars is shown.}
\label{fig:ebex_err}
\end{figure}

Next we study the power spectrum errors which are obtained using these weight functions.
In Fig.~\ref{fig:ebex_err}, three sets of power spectrum errors are compared.
Considering the leftmost set of error bars first 
(i.e. pseudo-$C_\ell$ with counterterms and without the point source mask),
we would like to know, how do these errors compare to optimal?
As previously remarked, we are unable to make this comparison directly since the fiducial experiment lacks azimuthal symmetry.
However, the pseudo-$C_\ell$ errors for the fiducial realistic experiment can be compared to the pseudo-$C_\ell$ errors
which were obtained previously (Fig.~\ref{fig:stcerr}) for an azimuthally symmetric mock survey with roughly the same sky coverage
and noise level.  We find that the two are roughly equal; the RMS bandpower errors for the fiducial realistic experiment
for both EE and BB are $\sim 10\%$ worse than the mock survey at low $\ell$ and $\sim 10\%$ better
at high $\ell$.  It seems plausible that this difference is simply due to inhomogeneity in
the noise, and that our method remains near-optimal for the fiducial realistic experiment, although we cannot
say this with absolute certainty since direct computation of the optimal errors is not feasible.

Now let us compare the errors with and without the point source mask, for pseudo-$C_\ell$ estimators with counterterms
(i.e., the left and middle error bars in Fig.~\ref{fig:ebex_err}).
We find that the point source mask degrades the errors on BB by $\sim 20\%$ for $\ell \simle 250$
but the effect becomes small for large $\ell$, or for EE on all angular scales.
Even though a tiny fraction of sky area is masked, we find that the impact on $B$-mode errors is not 
negligible.  This is to be expected since the point source mask does decrease the number of pure 
$B$-modes in the survey region.
However, even taking a conservative estimate for the number density of point sources,
we have found that this effect is not large.

Finally, we compare pseudo-$C_\ell$ with counterterms to ordinary pseudo-$C_\ell$
(i.e., the middle and right error bars in Fig.~\ref{fig:ebex_err}).
It is seen that at this noise level, including counterterms in the estimator is essential for $E$-$B$ separation.
If ordinary pseudo-$C_\ell$ estimators are used instead, the
RMS $B$-mode errors are worse by a factor $\sim$2-3 at intermediate to high $\ell$, and a factor
$\sim 10$ at low $\ell$.
If this comparison is repeated without the point source mask, we find that the two become equal
at high $\ell$ but remain drastically different at low $\ell$.
This suggests that the suboptimality of ordinary pseudo-$C_\ell$ is mainly due to $E\rightarrow B$ mixing
from survey boundaries at low $\ell$ and point sources at high $\ell$.

\section{Discussion}

We have given a general treatment of $E\rightarrow B$ mixing and argued that when estimating $B$-mode power,
no information is lost by passing to the curl $\chi_B$.
The analogue for a scalar field, passing to the Laplacian, would lose the largest-scale modes on a cut sky,
but for a spin-2 field it is the ambiguous modes which are lost in this way.
This presents a general approach to constructing estimators which are pure, i.e. which filter out
contributions from ambiguous modes.
We considered the estimator defined by applying pseudo-$C_\ell$ estimation to the scalar field $\chi_B$, and
proved that it is mathematically equivalent to the pure pseudo-$C_\ell$ estimator from \cite{Smith:2005gi}.
The equivalence sheds light on properties of the estimator, e.g. the requirement of a smoothly apodized weight function
can be understood as arising from colored noise in $\chi_B$.
Perhaps most importantly, it allows us to consider ordinary and pure pseudo-$C_\ell$ estimators in a unified
way when studying the problem of optimizing the weight function.

The central result of this paper is a general ansatz for optimizing the pseudo-$C_\ell$
weight function.
The ansatz can be expressed either in matrix form (Eq.~(\ref{eq:Wopt})), or as a variational principle:
the optimized weight function minimizes the total expectation value $\eE$ of the estimator,
subject to the constraint $\sum_x W(x)=1$.
Although our emphasis has been on next-generation ground-based polarization experiments,
and we have only considered noise which is uncorrelated between pixels (Eq.~(\ref{eq:uncorr})),
the ansatz is much more general: in principle it applies to arbitrary signal and noise covariance.
In particular, it should be useful for the temperature power spectrum, but we have not 
pursued this here.

\begin{table}
\begin{center}
\begin{tabular}{|c|c|c|}
\hline                          &    $(T/S)_{1\sigma}$ &  $\sigma(A_{\rm lens})$  \\  \hline
Mode-counting estimate (Eq.~(\ref{eq:modecount}))                                                      &  0.0032   &  0.095     \\  \hline
Homogeneous circular mock survey, optimal estimators (Fig.~\ref{fig:stcerr})                          &  0.0039   &  0.095     \\
Homogeneous circular mock survey, pseudo-$C_\ell$ with counterterms (Fig.~\ref{fig:stcerr})           &  0.0043   &  0.096     \\
``Realistic'' noise map, no point source mask, pseudo-$C_\ell$ with counterterms (Fig.~\ref{fig:ebex_err})      &  0.0045   &  0.082     \\
``Realistic'' noise map, point source mask applied, pseudo-$C_\ell$ with counterterms (Fig.~\ref{fig:ebex_err}) &  0.0056   &  0.085     \\  \hline
Ordinary pseudo-$C_\ell$                                                                               &  0.0536   &  0.258     \\  \hline
\end{tabular}
\end{center}
\caption{Forecasts for the $1\sigma$ upper limit on $(T/S)$ and the fractional error on the amplitude $A_{\rm lens}$
of the lensing $B$-mode, for various levels of approximation to the fiducial realistic experiment.}
\label{tab:ebexsumm}
\end{table}

We have shown our the ansatz can be solved numerically for an arbitrary spatial noise
distribution, and studied the performance of our estimators for several mock surveys,
culminating in a full simulation of a realistic experiment with characteristics based on the upcoming balloon borne EBEX:
complicated boundaries, inhomogeneous noise and a randomly generated point source mask (Fig.~\ref{fig:ebex_hitcount}).
This is summarized in Tab.~\ref{tab:ebexsumm}, where we have compressed the $B$-mode power spectrum
errors obtained for each survey into two numbers: the $1\sigma$ upper limit on $(T/S)$ and
the fractional error $\sigma(A_{\rm lens})$ on the overall amplitude of the lensing $B$-mode.
In computing $\sigma(A_{\rm lens})$, we have used the WMAP3 best-fit model \cite{Spergel:2006hy} with $\sigma_8=0.74$;
in a different fiducial model it would scale roughly as $\sigma(A_{\rm lens}) \propto \sigma_8^{-1}$.
(We have also treated the power spectrum covariance as Gaussian; including non-Gaussianity 
in the lensing $B$-mode is expected to give a $\sim 10\%$ correction to $\sigma(A_{\rm lens})$ 
at noise levels of our fiducial experiment \cite{Smith:2006nk} but the effect can become large for lower noise.)

As a baseline, we have shown in the first row of the table the naive ``mode-counting'' estimate,
given by assigning uncorrelated errors to each bandpower as follows:
\be
\Cov(\Delta_b,\Delta_b') = \frac{1}{2} \fsky \sum_{\ell\in b} (2\ell+1) 
                           \left( \frac{C_\ell^{BB}}{C_\ell^{BB} + N_\ell} \right)^2 \langle \Delta_b \rangle^2 \delta_{bb'}   \label{eq:modecount}
\ee
where $N_\ell$ is the noise power spectrum.
In the last row, we have shown for comparison the result of using ordinary pseudo-$C_\ell$ estimators
without counterterms.

Each row of the table isolates one effect which might be worrying for $E$-$B$ separation:
mixing from the survey boundary, suboptimality of pseudo-$C_\ell$, inhomogeneous noise with 
small-scale features, and the point source mask.
Our conclusion is that each of these has a small impact using our method,
although it is worth noting that when combined, these effects do
result in a value of $(T/S)_{1\sigma}$ which differs from the baseline
mode-counting estimate by 75\%.
Considering all of these in combination, we have given a complete treatment
of ``geometric'' effects which arise from the mask and inhomogeneous noise coupling $E$ and $B$ in
a realistic survey.

It should be mentioned that there are other, non-geometric effects not considered in this paper which will
have an impact in real CMB polarization experiments, such as $(1/f)$ noise, removal of ground-synchronous modes,
systematic errors, and foreground contamination.
In particular, the power spectrum errors we have presented in \S\ref{sec:ebex} should not be regarded as a
``bottom line'' forecast for EBEX, as these effects have not been included.
However, geometric mixing of $E$ and $B$ has been a practical concern for next generation experiments;
we have given a general solution to the problem and demonstrated the method for noise distributions with the
complexity of a real experiment.
Moreover, we have shown how to generate optimized weight functions in a completely automated way,
thus removing an outstanding practical obstacle in the pseudo-$C_\ell$ method.

\section*{Acknowledgements}
We thank Carlo Baccigalupi, Shaul Hanany, Wayne Hu, Max Tegmark and Bruce Winstein for useful discussions 
and Sam Leach and Will Grainger for providing the EBEX simulations. 
KMS was supported by the Kavli Institute for Cosmological Physics through the grant NSF PHY-0114422. 
M.~Z.~is supported by the Packard and Sloan foundations, NSF AST-0506556 and NASA NNG05GG84G.


\appendix

\section{Taking derivatives by finite differencing}
\label{finite}

In \S\ref{sec:purecl}, we proved that the pure pseudo-$C_\ell$ method is equivalent to computing the curl $\chi_B$ and then estimating
the power spectrum of the resulting scalar field.
Throughout this paper, we have mainly used the former approach, but have also found that the latter can also be used as a practical technique
when analyzing noisy maps.
Here, we describe some implementational details of the method; we will see that this also sheds light on the pixel resolution which is needed 
to control $E\rightarrow B$ aliasing.

Making a $\chi_B$ map out of polarization data involves taking finite differences. 
The derivative is a local operation so to understand the requirements it is easier to work in the flat sky approximation. For this analysis to apply, the flat sky approximation should be valid over the scale of a pixel, which is almost always true. 

In this case we have:  
\beqa
\chi_E&=& (\partial^2_x-\partial^2_y) Q + 2 \partial^2_{xy} U  \nonumber \\
\chi_B&=& (\partial^2_x-\partial^2_y) U - 2 \partial^2_{xy} Q.
\eeqa
In the discrete case we take the partial derivatives by doing linear combinations of  neighboring pixels. To isolate $B$ around a point $\r$ we consider a combination,
\beq
\hat \chi_{B} = \sum_p \W_p^T\cdot \P_p,
\eeq
where the sum over $p$ is a sum over neighboring pixels and $\P=\left(Q, U\right)$. The strategy to determine the weights is to consider all Fourier modes, one at a time,  and try to reduce the $E$ leakage they produce.  
We demand that for the mode of wavevector $\k$:  
\beq
\hat \chi_{B} = -k^2  e^{i\k\cdot\r} [ B(\k) (1 + a_B(\hat \k) (k \Delta r)^p \cdots)  +  E(\hat \k) (a_E(\k) (k \Delta r)^q +\cdots ) ],
\label{contamin}
\eeq
where $\Delta r$ measures the separation to the closest pixel and $a_{E,B}$ are order one coefficients which depend on the orientation of $\k$ relative to the grid. 
The aim is to make $p$ and $q$ as large as possible. The weights of course are independent of $\k$ but will depend on the shape of the grid around the point being considered. 
The grid will also determine the coefficients $a_{E,B}$.

For a square grid one can obtain the following results: using the 8 neighboring pixels (the three by three square with the point of interest at the center) 
one can get $(p,q)=(2,2)$, using 16 one can get $(p,q)=(4,4)$ and with 24 neighbors (a 5 by 5 square centered on the pixel) one gets $(p,q)=(2,6)$. 
Due to the symmetries of the grid one can parametrize the weights in the above cases by up to six coefficients, $w_1,\cdots, w_6$. 
For 8 neighbors only $w_1$ and $w_2$ are needed, for 16 $w_1$ through $w_4$, and all of them are used for the 24 neighbor case. 
The required linear combinations are: 
\beqa
\hat \chi_{B,(i,j)}&=& w_1 (u_{i-1,j}+u_{i,j-1}-u_{i+1,j}-u_{i,j+1}) \nonumber \\
&+& w_2 (q_{i+1,j+1}+q_{i-1,j-1}-q_{i-1,j+1}-q_{i+1,j-1}) \nonumber \\
&+& w_3 (u_{i-2,j}+u_{i,j-2}-u_{i+2,j}-u_{i,j+2}) \nonumber \\
&+& w_4 (q_{i+2,j+2}+q_{i-2,j-2}-q_{i-2,j+2}-q_{i+2,j-2}) \nonumber \\
&+& w_5 ( u_{i+2,j-1}+u_{i+1,j-2}+u_{i-1,j+2}+u_{i-2,j+1} \nonumber \\ &-&  \qquad u_{i+2,j+1}-u_{i+1,j+2}-u_{i-1,j-2}-u_{i-2,j-1}) \nonumber \\
&+& w_6 (q_{i+2,j+1}+q_{i+2,j-1}+q_{i-2,j+1}+q_{i-2,j-1} \nonumber \\ &-&  \qquad q_{i+1,j+2}-q_{i-1,j-2}-q_{i+1,j-2}-q_{i-1,j+2})
\label{chib}
\eeqa
with $w_i$ shown in table \ref{tablew}. The coefficients for $\chi_E$ can be obtained from the above by simply rotating the polarization by 45 degrees. 

For the lowest order the weights are exactly what one expects for those second derivatives. 
As one increases the number of neighbors the order up to which one can suppress the leakage increases. To first and second orders the answers we present are unique. For the third order case the requirement to have $(p,q)=(2,6)$ does not determine the weights but leaves one degree of freedom. Than freedom is not enough to increase either $p$ or $q$ so we made the choice that minimized $\langle a_B^2\rangle$.

\begin{figure}
\centerline{\epsfxsize=4.0truein\epsffile{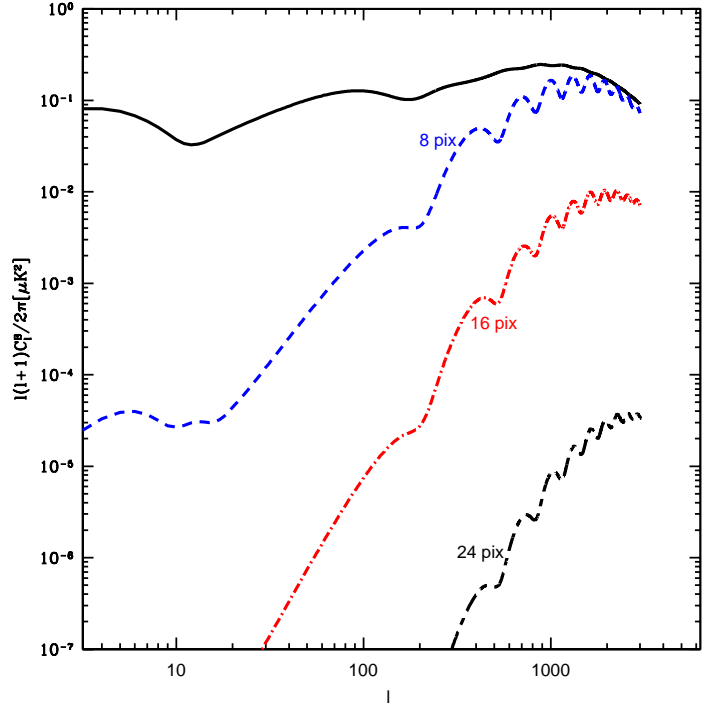}}
\caption{Expected $B$-mode power spectrum including the lensing signal and GW with $T/S=0.1$ compared to the level of leakage for three different numbers of neighbors. For the leakage we plot   power spectrum and $\chi_B$ in the absence of any $B$-mode signal times  $(l-2)! /(l+2)!$. If $\Delta r$ gives the separation between neighboring pixels in the grid, the successive approximations to the derivatives scale as $\Delta r^2$, $\Delta r^4$ and $\Delta r^6$. }
\label{derivatives}
\end{figure}

Figure \ref{derivatives} shows the leakage power spectra for a square grid with 1 arcminute  pixel separation. What is plotted is the $\chi_B$ power spectrum times $(l-2)! /(l+2)!$ in the absence of cosmological $B$-modes.  It is clear that in this case using just the nearest neighbors to take the derivatives is sufficient, except perhaps at relatively large $l$ for the lensing signal. In that case an extra order might be required. Note for example that for an experiment like EBEX the leakage power is well below the noise for all relevant scales.  It is also apparent that if one is after the gravity wave signal the requirements are significantly relaxed. In fact for $T/S=0.01$ and lowest order derivative, the leakage power spectrum is a factor of ten below the gravity wave signal al $l=100$ even for pixel separations of $10$ arcminutes. 

\begin{table}[htdp]
\caption{Weights used in equation (\ref{chib}) to calculate $\hat \chi_B$ for three different choices for the number of neighbors used. The last four columns quantify the level of residual contamination using the quantities in equation (\ref{contamin}). }
\begin{center}
\begin{tabular}{||c|c|c|c|c|c|c|c|c|c|c||}
\hline
  & $w_1$ & $w_2$ & $w_3$ & $w_4$& $w_5$& $w_6$ & p & q& $\langle a_E^2 \rangle$ & $ \langle a_B ^2\rangle$  \\
\hline
8 & 1 & $1\over 2$ & 0 & 0 & 0 & 0 & 2 & 2& $1.6 \times 10^{-2}$ & $8.7 \times 10^{-3}$  \\
\hline
16 & $4\over 3$ & $2\over 3$ & -$1\over 12$ & -$1\over 24$ & 0  & 0 & 4& 4& $8.2 \times 10^{-4}$  & $1.1 \times 10^{-4}$ \\
\hline
24 &$806 \over 2625$ & $2243\over 7875$ & $1501 \over 10500$ &-$ 239\over 63000$ &$53\over 2625$& $907\over 15750$ & 2 & 6 &  $7.1 \times 10^{-2}$ & $3.2 \times 10^{-9}$ \\
\hline
\end{tabular}
\end{center}
\label{tablew}
\end{table}%

\section{Relation to FKP}
\label{sec:FKP}

In this appendix we prove the assertion, stated without proof at the end of \S\ref{sec:conjecture}, that our ansatz for the
optimal weight function reduces to the FKP \cite{Feldman:1993ky} approximation under appropriate conditions. 
The FKP approximation is used to incorporate the effect of inhomogeous noise when estimating the amplitude of modes that are 
much shorter than the size of the survey and also of the typical variation scale of the noise.  FKP propose a weight 
\beq
W^{FKP}_i \propto \frac{1}{\sigma^2_i + \Omega^{-1} C_{l_0}},
\eeq
where $l_0$ is the wavenumber where the power spectrum is being estimated,
$\Omega$ is the solid angle of each pixel, and $\sigma^2_i$ the noise variance in pixel $i$.
Working in the flat sky approximation, we will now show that under the FKP assumptions, $W^{FKP}_i$ satisfies Eq.~(\ref{ansatzeq}).

Assuming that we are estimating power in a narrow band $\alpha$ near $l_0$ we have,
\beqa
\bC_{ij}  &=& \sigma^2_i \delta_{ij} + \int {d^2\l \over (2\pi)^2} C_l e^{-i\l\cdot (\r_i-\r_j)}. \\
\bCa_{ij} &=& \int \frac{d\varphi_{\l_0}}{2\pi} e^{-i\l_0\cdot (\r_i-\r_j)} \nonumber 
\eeqa
where $\l_0 = l_0(\cos\varphi_{\l_0},\sin\varphi_{\l_0})$.
Plugging this into Eq.~(\ref{ansatzeq})  we get,
\beq
\sum_j \bC_{ij} \bCa_{ij} W_j^{FKP} =
\sum_j \left( \sigma^2_i \delta_{ij} + \int {d^2\l \over (2\pi)^2} C_l e^{-i\l\cdot (\r_i-\r_j)} \right)
\left( \int \frac{d\varphi_{\l_0}}{2\pi} e^{-i\l_0\cdot (\r_i-\r_j)} \right)
\frac{1}{\sigma^2_j + \Omega^{-1} C_{l_0}}  \label{eq:FKP1}
\eeq
The first term,  coming from the noise contribution to the covariance, simply gives:
\beq
\frac{\sigma^2_i}{\sigma^2_i + \Omega^{-1} C_{l_0}}.
\label{neq}
\eeq
To make progress on the second term we need to use some of the FKP assumptions. 
First we will use that we are interested in a situation in which the noise varies slowly and write,
\beq
\frac{1}{\sigma^2_j + \Omega^{-1} C_{l_0}} \equiv f(\r_j) = f(\r_i) + \nabla f  (\r_i) (\r_j - \r_i)+\cdots ,
\eeq
so that the second term in~(\ref{eq:FKP1}) becomes
\beq
\sum_j \int \frac{d^2\l}{(2\pi)^2} \frac{d\varphi_{\l_0}}{2\pi}\,
C_l e^{-i(\l+\l_0)\cdot (\r_i-\r_j)}
\Big(  f(\r_i) +  \nabla f  (\r_i) (\r_j - \r_i)+\cdots \Big)
\label{feq}
\eeq
For the first of these terms we get:
\ba
\int \frac{d^2\l}{(2\pi)^2} \frac{d\varphi_{\l_0}}{2\pi}\,
C_l f(r_i) \sum_j e^{-i(\l+\l_0)\cdot (\r_i-\r_j)}
&=& 
\int \frac{d^2\l}{(2\pi)^2} \frac{d\varphi_{\l_0}}{2\pi}\,
C_l f(r_i) \Omega^{-1} \Big[ (2\pi)^2 \delta^2(\l+\l_0) + \bigoh(1/l_0 R) \Big]   \nn \\
&=&
\frac{\Omega^{-1} C_{l_0}}{\sigma_i^2 + \Omega^{-1}C_{l_0}} + \bigoh(1/l_0 R)
\ea
where $R$ is the size scale of the patch.
Now let us check what happens with the additional terms in Eq.~(\ref{feq}) describing anisotropy of the noise.
The additional contribution is proportional to: 
\beq
\nabla_{\r} f \cdot \nabla_{\l_0} C_{l_0},
\eeq
which is suppressed by $1/l_0 R_N$ with respect to the leading term where $R_N$ gives the scale of variation of the noise. 

Putting all of this together, we see that
\be
\sum_j \bC_{ij} \bCa_{ij} W_j^{FKP} = 1 + \bigoh(1/l_0 R) + \bigoh(1/l_0 R_N)
\ee
i.e., for diagonal inhomogenous noise when considering modes much shorter than both the scale of variation of the noise and the patch, 
our ansatz~(\ref{eq:Wopt}) for the pseudo-$C_\ell$ weight function reduces to FKP.

\section{Numerical optimization of weight functions}
\label{app:cg}

In this appendix we supply the details of our procedure, described briefly
in \S\ref{sec:numerical}, for numerically optimizing pseudo-$C_\ell$ weight functions 
given the noise RMS $\sigma(x)$ in each pixel.
We first treat the more difficult case of optimizing the $B$-mode weight function;
as we will then see, optimizing the $E$-mode weight function can be obtained as
a special case.
As described in \S\ref{sec:numerical}, the $B$-mode ``weight function'' really consists of
three functions $W(x)$, $W_1(x)$, $W_2(x)$ with spins 0, 1, 2 respectively.
These are varied independently in order to minimize the total expectation value
$\eE$ of the pseudo-$C_\ell$ estimator (with contributions from signal and noise), 
subject to the normalization constraint
\be
\sum_x W(x) = 1\, .  \label{eq:appnc}
\ee

To solve this minimization problem, we first rewrite the expectation value $\eE$ to be minimized
in a form which makes the dependence on the weight functions easier to understand.
Considering first the noise contribution to $\eE$, a short calculation shows that 
\be
\eEn = \frac{1}{4\pi} \sum_{\ell x} W_\ell\,\, \sigma^2(x)
\left( W(x)^2 + 4 \frac{|W_1(x)|^2}{(\ell-1)(\ell+2)} + \frac{|W_2(x)|^2}{(\ell-1)\ell(\ell+1)(\ell+2)} \right)  \label{eq:een}
\ee
where $W_\ell$ denotes the $\ell$ weighting within the bandpower under consideration.
(Throughout this paper we have taken this to be $W_\ell = \ell(\ell+1)/(2\pi)$ inside the
band and zero outside the band, as in Eq.~(\ref{eq:pseudoC}).)

The signal contribution to $\eE$ is more difficult to compute, but follows from the results
of Appendix~D in \cite{Smith:2005gi}.
In order to write it down, we decompose the pieces of the weight function in spherical harmonics:
\ba
a^W_{\ell m} &=& \sum_x W(x) Y^*_{\ell m}(x)  \\
a^G_{\ell m} = -\sum_x W_1(x) \left( \frac{{}_1Y_{\ell m}^*(x) - {}_{-1}Y_{\ell m}^*(x)}{2} \right)  & \qquad & 
a^C_{\ell m} = i \sum_x W_1(x) \left( \frac{{}_1Y_{\ell m}^*(x) + {}_{-1}Y_{\ell m}^*(x)}{2} \right)  \label{eq:wtgc}  \\
a^E_{\ell m} = -\sum_x W_2(x) \left( \frac{{}_2Y_{\ell m}^*(x) + {}_{-2}Y_{\ell m}^*(x)}{2} \right)  & \qquad & 
a^B_{\ell m} = i \sum_x W_2(x) \left( \frac{{}_2Y_{\ell m}^*(x) - {}_{-2}Y_{\ell m}^*(x)}{2} \right)  \label{eq:wteb}
\ea
(The spin-2 piece~(\ref{eq:wteb}) is the E/B decomposition from Eq.~(\ref{eb}), whereas the spin-1 piece~(\ref{eq:wtgc}) is
the gradient/curl decomposition of a vector field.)

The signal contribution to $\eE$ can then be written in the following form:
\be
\eEs = \sum_{\ell m}
\left( \begin{array}{ccc} a^{W*}_{\ell m} & a^{G*}_{\ell m} & a^{E*}_{\ell m} \end{array} \right)
\left( \begin{array}{ccc}
           C_\ell^{WW}  &  C_\ell^{WG}  &   C_\ell^{WE}  \\
           C_\ell^{WG}  &  C_\ell^{GG}  &   C_\ell^{GE}  \\
           C_\ell^{WE}  &  C_\ell^{GE}  &   C_\ell^{EE}
\end{array} \right)
\left( \begin{array}{c} a^W_{\ell m} \\ a^G_{\ell m} \\ a^E_{\ell m} \end{array} \right)
+
\left( \begin{array}{cc} a^{C*}_{\ell m} & a^{B*}_{\ell m} \end{array} \right)
\left( \begin{array}{ccc}
           C_\ell^{CC}  &  C_\ell^{CB} \\
           C_\ell^{CB}  &  C_\ell^{BB}
\end{array} \right)
\left( \begin{array}{c} a^C_{\ell m} \\ a^B_{\ell m} \end{array} \right)   \label{eq:ees}
\ee
Here, the power spectra which appear depend on the bandpower weighting $W_\ell$ and signal power spectra.
For the latter, it is convenient to introduce the notation $C^\pm_\ell = (C^{EE,\rm sig}_\ell \pm C^{BB,\rm sig}_\ell)$.
The power spectra in Eq.~(\ref{eq:ees}) are then given by:
\ba
C_\ell^{WW} &=& \int_{-1}^1 dz\, \sum_{\ell'\ell''} \left( \frac{2\ell''+1}{16\pi} \right) W_{\ell''}
                  d_{00}^\ell(z)  \Big[ C_{\ell'}^+ d_{22}^{\ell'}(z) d_{22}^{\ell''}(z) -
                       C_{\ell'}^- d_{2,-2}^{\ell'}(z) d_{2,-2}^{\ell''}(z) \Big]  \label{eq:appcl} \\
C_\ell^{WG} &=& \int_{-1}^1 dz\, \sum_{\ell'\ell''} \left( \frac{2\ell''+1}{16\pi} \right) 
                     \frac{2 W_{\ell''}}{\sqrt{(\ell''-1)(\ell''+2)}}
                  d_{01}^\ell(z)  \Big[-C_{\ell'}^+ d_{22}^{\ell'}(z) d_{12}^{\ell''}(z) +
                       C_{\ell'}^- d_{2,-2}^{\ell'}(z) d_{1,-2}^{\ell''}(z) \Big]   \nn  \\
C_\ell^{WE} &=& \int_{-1}^1 dz\, \sum_{\ell'\ell''} \left( \frac{2\ell''+1}{16\pi} \right) 
                     \frac{W_{\ell''}}{\sqrt{(\ell''-1)\ell''(\ell''+1)(\ell''+2)}}
                  d_{02}^\ell(z)  \Big[ -C_{\ell'}^+ d_{22}^{\ell'}(z) d_{02}^{\ell''}(z) +
                       C_{\ell'}^- d_{2,-2}^{\ell'}(z) d_{02}^{\ell''}(z) \Big]     \nn  \\
C_\ell^{GG} &=& \int_{-1}^1 dz\, \sum_{\ell'\ell''} \left( \frac{2\ell''+1}{16\pi} \right) 
                     \frac{4 W_{\ell''}}{(\ell''-1)(\ell''+2)}
                     \Big[ C_{\ell'}^+ d_{11}^\ell(z) d_{22}^{\ell'}(z) d_{11}^{\ell''}(z) -
                           C_{\ell'}^- d_{1,-1}^\ell(z) d_{2,-2}^{\ell'}(z) d_{1,-1}^{\ell''}(z) \Big]  \nn \\
C_\ell^{CC} &=& \int_{-1}^1 dz\, \sum_{\ell'\ell''} \left( \frac{2\ell''+1}{16\pi} \right) 
                     \frac{4 W_{\ell''}}{(\ell''-1)(\ell''+2)}
                     \Big[ C_{\ell'}^+ d_{11}^\ell(z) d_{22}^{\ell'}(z) d_{11}^{\ell''}(z) +
                           C_{\ell'}^- d_{1,-1}^\ell(z) d_{2,-2}^{\ell'}(z) d_{1,-1}^{\ell''}(z) \Big]  \nn \\
C_\ell^{GE} &=& \int_{-1}^1 dz\, \sum_{\ell'\ell''} \left( \frac{2\ell''+1}{16\pi} \right) 
                     \frac{2 W_{\ell''}}{(\ell''-1)(\ell''+2) \sqrt{\ell''(\ell''+1)}}
                     \Big[ C_{\ell'}^+ d_{12}^\ell(z) d_{22}^{\ell'}(z)  -
                           C_{\ell'}^- d_{1,-2}^\ell(z) d_{2,-2}^{\ell'}(z)  \Big] d_{01}^{\ell''}(z) \nn  \\
C_\ell^{CB} &=& \int_{-1}^1 dz\, \sum_{\ell'\ell''} \left( \frac{2\ell''+1}{16\pi} \right) 
                     \frac{2 W_{\ell''}}{(\ell''-1)(\ell''+2) \sqrt{\ell''(\ell''+1)}}
                     \Big[ C_{\ell'}^+ d_{12}^\ell(z) d_{22}^{\ell'}(z) +
                           C_{\ell'}^- d_{1,-2}^\ell(z) d_{2,-2}^{\ell'}(z) \Big] d_{01}^{\ell''}(z)  \nn  \\
C_\ell^{EE} &=& \int_{-1}^1 dz\, \sum_{\ell'\ell''} \left( \frac{2\ell''+1}{16\pi} \right) 
                     \frac{W_{\ell''}}{(\ell''-1)\ell''(\ell''+1)(\ell''+2)}
                     \Big[ C_{\ell'}^+ d_{22}^\ell(z) d_{22}^{\ell'}(z) -
                           C_{\ell'}^- d_{2,-2}^\ell(z) d_{2,-2}^{\ell'}(z) \Big] d_{00}^{\ell''}(z)  \nn  \\
C_\ell^{BB} &=& \int_{-1}^1 dz\, \sum_{\ell'\ell''} \left( \frac{2\ell''+1}{16\pi} \right) 
                     \frac{W_{\ell''}}{(\ell''-1)\ell''(\ell''+1)(\ell''+2)}
                     \Big[ C_{\ell'}^+ d_{22}^\ell(z) d_{22}^{\ell'}(z) +
                           C_{\ell'}^- d_{2,-2}^\ell(z) d_{2,-2}^{\ell'}(z)  \Big] d_{00}^{\ell''}(z)  \nn
\ea
where $d_{ss'}^\ell(z)$ is the reduced Wigner rotation matrix element \cite{VMK}; note
that $d_{00}^\ell(z)$ is just the Legendre polynomial $P_\ell(z)$.
We have written the power spectra as position-space integrals following \cite{Smith:2005gi};
if convenient these expressions can be converted to harmonic space using the identity
\be
\int_{-1}^1 dz\, d_{s,\pm s'}^\ell(z) d_{2,\pm 2}^{\ell'}(z) d_{2-s,\pm 2 \mp s'}^{\ell''}(z)
= 2 \threej{\ell}{\ell'}{\ell''}{s}{-2}{2-s} \threej{\ell}{\ell'}{\ell''}{s'}{\mp 2}{\pm 2 \mp s'}
\ee

Armed with the expressions~(\ref{eq:een}) and~(\ref{eq:ees}) for the noise and signal contributions to $\eE$,
a procedure for minimizing $\eE = \eEs + \eEn$ subject to the constraint~(\ref{eq:appnc}) is not difficult
to come by.  Representing the three components of the $B$-mode weight function by a single length-$(5\npix)$
vector $w$, both $\eEs$ and $\eEn$ are quadratic in $w$, so we can write them compactly as
\be
\eE = \eEs + \eEn = w^T Q_s w + w^T Q_n w
\ee
where $Q_s$ and $Q_n$ are $(5\npix)$-by-$(5\npix)$ matrices.
In this notation, the constraint~(\ref{eq:appnc}) can be written $v^T w = 1$,
where $v$ is a vector with $\npix$ entries equal to 1 (corresponding to the spin-0 piece of
the weight function) and $(4\npix)$ entries equal to 0 (corresponding to the higher-spin pieces).
The solution to this minimization problem is simply
\be
w_{\rm opt} = \frac{(Q_s + Q_n)^{-1} v}{v^T (Q_s + Q_n)^{-1} v}  \label{eq:appsol}
\ee
Since dense inversion of a $(5\npix)$-by-$(5\npix)$ matrix is computationally infeasible,
our approach is to compute $(Q_s+Q_n)^{-1}{\bf w}$ using conjugate gradient inversion~\cite{NR}.

In order to use conjugate gradient inversion, two ingredients are needed.
First, one needs an efficient way to multiply a vector by $(Q_s+Q_n)$,
even though this matrix will be too large to store in dense form.
This is easily obtained from Eqs.~(\ref{eq:ees}) and~(\ref{eq:een}); note that multiplying by
$Q_s$ requires transforming from pixel to harmonic space and back.
The second ingredient is a preconditioner, or approximate inverse of 
$(Q_s+Q_n)$, used to speed convergence; the better the preconditioner
approximates $(Q_s+Q_n)^{-1}$, the more rapidly the conjugate gradient
search will converge.

We use the following simple preconditioner, obtained by
keeping only the diagonal of $(Q_s+Q_n)$, then inverting:
\be
\Big( W_0(x) \quad\quad W_1(x) \quad\quad W_2(x) \Big) \longrightarrow
\left( \frac{W_0(x)}{\sigma_0^2 + \sigma^2(x)} \quad\quad
                          \frac{W_1(x)}{\sigma_1^2 + \sigma^2(x)} \quad\quad
                          \frac{W_2(x)}{\sigma_2^2 + \sigma^2(x)}
\right)  \label{eq:apppc}
\ee
where we have defined
\ba
\sigma_0^2 &=& \sum_\ell \left( \frac{2\ell+1}{4\pi} \right) C_\ell^{WW}  \\
\sigma_1^2 &=& \sum_\ell \left( \frac{2\ell+1}{4\pi} \right) (C_\ell^{GG} + C_\ell^{CC})   \nn \\
\sigma_2^2 &=& \sum_\ell \left( \frac{2\ell+1}{4\pi} \right) (C_\ell^{EE} + C_\ell^{BB})\, . \nn
\ea
The performance of this preconditioner depends mainly on the noise level.
For noise levels $\simge 20$ $\mu$K-arcmin, only a few CG iterations are needed; for the
noise levels of our fiducial experiment ($\sim 5.75$ $\mu$K-arcmin), around 100 iterations are needed; and for noise
levels $\simle 1$ $\mu$K-arcmin, the CG search does not converge.
A better preconditioner may accelerate weight function optimization,
but we have not attempted to find one in this paper.

In summary, we have now arrived at a method for numerically optimizing the $B$-mode
weight function, given the bandpower (specified by the $\ell$ weighting $W_\ell$) and noise
RMS $\sigma(x)$.
One precomputes the power spectra in Eqs.~(\ref{eq:appcl}), then computes $(Q_s+Q_n)^{-1}v$
by conjugate gradient inversion, using the preconditioner~(\ref{eq:apppc}), to obtain the optimized
weight function given by~(\ref{eq:appsol}).

Finally, let us discuss the completely analagous but simpler problem of optimizing the $E$-mode weight function.
In this case, the weight function is just a scalar $W(x)$.  The noise and signal contributions
to $\eE$ are given (in analogy to Eqs.~(\ref{eq:een}),~(\ref{eq:ees}) above) by
\ba
\eEn &=& \frac{1}{4\pi} \sum_{\ell x} W_\ell\,\, \sigma^2(x) W(x)^2  \\
\eEs &=& \sum_{\ell m} a^{W*}_{\ell m} C_\ell^{WW} a^W_{\ell m}      
\ea
where
\be
C_\ell^{WW} \eqdef \int_{-1}^1 dz\, \sum_{\ell'\ell''} \left( \frac{2\ell''+1}{16\pi} \right) W_{\ell''}
                  d_{00}^\ell(z)  \Big[ C_{\ell'}^+ d_{22}^{\ell'}(z) d_{22}^{\ell''}(z) +
                       C_{\ell'}^- d_{2,-2}^{\ell'}(z) d_{2,-2}^{\ell''}(z) \Big]
\ee
Interpreting these as an $\npix$-by-$\npix$ matrix equation
\be
\eE = \eEn + \eEs = w^T Q_n w + w^T Q_s w
\ee
the optimized $E$-mode weight function is given by
\be
w_{\rm opt} = \frac{(Q_s+Q_n)^{-1}v}{v^T(Q_s+Q_n)^{-1}v}\,.  \label{eq:appsole}
\ee
where $v$ is a length-$\npix$ vector consisting of all 1's.
Eq.~(\ref{eq:appsole}) is evaluated using conjugate gradient inversion
with preconditioner given by  $W(x) \rightarrow W(x)/(\sigma_0^2 + \sigma(x)^2)$.

\end{document}